\documentclass[sigconf,screen]{acmart}

\usepackage[inline]{enumitem}
\usepackage[skip=0pt]{caption}
\usepackage{multirow}
\usepackage{xspace}
\usepackage{acronym}
\usepackage{graphicx}   % for \resizebox
\usepackage{booktabs}   % for \toprule/\midrule/\bottomrule
\usepackage{multirow}
\usepackage{graphicx}
\usepackage{makecell}
\usepackage{stfloats}
\usepackage{float}
\usepackage{placeins}
\usepackage{adjustbox}
\usepackage{tikz}
\usepackage{pgfplots}
\usepackage[table]{xcolor}
\usepackage[dvipsnames]{xcolor}
\pgfplotsset{compat=1.17}
\usetikzlibrary{patterns}
\usepackage{subcaption}
\usetikzlibrary{decorations.pathreplacing}
\renewcommand\footnotetextcopyrightpermission[1]{}

\AtBeginDocument{%
  \providecommand\BibTeX{{%
    \normalfont B\kern-0.5em{\scshape i\kern-0.25em b}\kern-0.8em\TeX}}}
\makeatletter
\newcommand\headingnodot{\def\@toclevel{4}%
  \@startsection{paragraph}{4}{\z@}%
  {-.2\baselineskip \@plus -2\p@ \@minus -.2\p@}%
  {-3.5\p@}%
  {\ACM@NRadjust{\bfseries}}}
\newcommand{\heading}[1]{\headingnodot{#1.}}
\makeatother

\acrodef{QA}{question answering}
\acrodef{GaQR}{generation-augmented question rewriter}
\acrodef{LLM}{large language model}
\acrodef{RAG}{retrieval-augmented generation}
\acrodef{QU}{query understanding}
\acrodef{IR}{information retrieval}
\acrodef{PRF}{pseudo‑relevance feedback}
\acrodef{GRF}{generative-relevance feedback}
\acrodef{CoT}{chain-of-thought}
\acrodef{KLD}{Kullback-Leibler divergence}
\acrodef{BGAE}{Balanced Generation-Augmented Expansion}
\acrodef{GRPO}{Group Relative Policy Optimization}

\newcommand{\sigp}{\rlap{$^{+}$}}
\newcommand{\sigm}{\rlap{$^{-}$}}
\newcommand{\num}[1]{$#1$}
\newcommand{\nump}[1]{$#1$\sigp}
\newcommand{\numm}[1]{$#1$\sigm}
\newcommand{\numb}[1]{$\mathbf{#1}$}
\newcommand{\numbp}[1]{$\mathbf{#1}$\sigp}

\newcommand{\numu}[1]{$\underline{#1}$}
\newcommand{\numup}[1]{$\underline{#1}$\sigp}
\newcommand{\numum}[1]{$\underline{#1}$\sigm}
\author{Yunfei Zhong}
\authornote{Co-first authors. Work done during internship at Baidu Inc.}
\affiliation{%
  \institution{Institute of Computing Technology, Chinese Academy of Sciences}
  \city{Beijing}
  \country{China}}
\email{zhongyunfei25@mails.ucas.ac.cn}

\author{Jun Yang}
\authornotemark[1] % 共用上一个authornote
\affiliation{%
  \institution{Institute of Computing Technology, Chinese Academy of Sciences}
  \city{Beijing}
  \country{China}}
\email{yangjun24s@ict.ac.cn}

\author{Yixing Fan}
\authornote{Corresponding author.}
\affiliation{%
  \institution{Institute of Computing Technology, Chinese Academy of Sciences}
  \city{Beijing}
  \country{China}}
\email{fanyixing@ict.ac.cn}

\author{Lixin Su}
\affiliation{%
  \institution{Baidu Inc.}
  \city{Beijing}
  \country{China}}
\email{sulixinict@gmail.com}

\author{Maarten de Rijke}
\affiliation{%
  \institution{University of Amsterdam}
  \city{Amsterdam}
  \country{The Netherlands}}
\email{m.derijke@uva.nl}

\author{Ruqing Zhang}
\affiliation{%
  \institution{Institute of Computing Technology, Chinese Academy of Sciences}
  \city{Beijing}
  \country{China}}
\email{zhangruqing@ict.ac.cn}

\author{Xueqi Cheng}
\affiliation{%
  \institution{Institute of Computing Technology, Chinese Academy of Sciences}
  \city{Beijing}
  \country{China}}
\email{cxq@ict.ac.cn}

\renewcommand{\shortauthors}{Y. Zhong et al.}

\begin{document}

\title[Reason to Retrieve: Enhancing Query Understanding through Decomposition and Interpretation]{Reason to Retrieve: Enhancing Query Understanding\\ through Decomposition and Interpretation}

\begin{abstract} 
\Ac{QU} aims to accurately infer user intent to improve document retrieval. It plays a vital role in modern search engines. While \acp{LLM} have made notable progress in this area, their effectiveness has primarily been studied on short, keyword-based queries. With the rise of AI-driven search, long-form queries with complex intent become increasingly common, but they are underexplored in the context of LLM-based \ac{QU}.
To address this gap, we introduce \textbf{ReDI}, a \textbf{reasoning}-enhanced query understanding method through \textbf{decomposition} and \textbf{interpretation}.
ReDI uses the reasoning and understanding capabilities of \acp{LLM} within a three-stage pipeline. (i) It decomposes a complex query into a set of targeted sub-queries to capture the user intent. (ii) It enriches each sub-query with detailed semantic interpretations to enhance the retrieval of intent-document matching. 
And (iii), after independently retrieving documents for each sub-query, ReDI uses a fusion strategy to aggregate the results and obtain the final ranking.
We collect a large-scale dataset of real-world complex queries from a commercial search engine and distill the query understanding capabilities of DeepSeek-R1 into small models for practical application.
Experiments on public benchmarks, including BRIGHT and BEIR, show that ReDI consistently outperforms strong baselines in both sparse and dense retrieval paradigms, demonstrating its effectiveness. We release our code, generated sub-queries, and interpretations at \url{https://github.com/youngbeauty250/ReDI}.
\end{abstract}

\maketitle
\acresetall
\begin{comment}

\end{comment}

\section{Introduction}
\Ac{QU} aims to infer the user’s intent behind a query to improve the retrieval of relevant documents. It has become a fundamental component of modern search engines~\cite{Yi20}, as it is both effective and straightforward to integrate into existing search systems. However, due to the inherent flexibility of language and the implicit nature of user intent, accurately inferring the user's true information needs from their query is a significant challenge.

Prior \ac{QU} work primarily uses \emph{knowledge-based} or \emph{pseudo-rele\-vance feedback} methods.
Knowledge-based methods~\cite{10.5555/188490.188508,10.1145/2513204.2513209,10.1145/988672.988763,10.1145/1718487.1718493} enrich query representations using structured resources like WordNet, Wikipedia, or user logs. 
In contrast, \ac{PRF}-based methods \cite{10.1145/383952.383972,10.1108/eb026866,10.1145/366836.366860} extract content from the top-$k$ retrieved documents, which are assumed to be relevant to the initial query. 
Despite performance gains, both types of method rely on fixed heuristics or initial retrieval quality. 
This limits their ability to capture latent intent and often causes query drift or misinterpretations with ambiguous or concise queries~\cite{10.1145/2071389.2071390}.

In recent years, \ac{LLM}-based \acl{QU} methods have proven effective by using linguistic and world knowledge learned during pre-training \cite{Gao2022PreciseZD,Wang2023Query2docQE}. These approaches prompt an \ac{LLM} to infer user intent and optimize it to capture richer semantic representations aligned with target documents \cite{Ma2023QueryRF,chan2024rq,zheng2024take}. Most prior studies have evaluated their effectiveness within traditional retrieval settings, where users issue keyword-based queries to locate relevant documents for a specific task, a process we term \textbf{information-locating retrieval}.

With the rapid advancement of LLM reasoning and generation capabilities, AI-driven search systems~\cite{openai2025deep,guo2025deepseek,gemini2025deep} have enabled more complex forms of information seeking. Here, user queries often involve multiple entities, extended temporal scopes, and diverse knowledge domains, demanding sophisticated reasoning. e.g., a query like \textit{``How has the relationship between scientific development and capital evolved from the Industrial Revolution to the present?''} requires the integration of dispersed historical evidence and conceptual synthesis. We refer to this paradigm as \textbf{reasoning-intensive retrieval}~\cite{su2025bright}, which poses substantial challenges for existing systems that struggle to accurately interpret, decompose, and satisfy such multifaceted information needs.

\vspace{-0.6em}
\heading{Research goal}
Recent studies have sought to exploit the reasoning capabilities of \acp{LLM} to enhance retrieval performance on complex tasks, e.g., ReasonIR~\cite{Shao2025ReasonIRTR}, ReasonRank~\cite{liu2025reasonrank}, and DIVER~\cite{Long2025DIVERAM}. While these approaches yield strong performance, they often come with substantial computational costs. 
A natural way to handle complexity is through divide-and-conquer strategies. However, prior work suggests that an information-rich long query often outperforms decomposed ones~\cite{Shao2025ReasonIRTR}. 
This raises the key question that we investigate: 
\textbf{when addressing complex queries, is query decomposition inherently ineffective, or is its application the real issue?}

\heading{Proposed \acl{QU} method}
To address our key question, we propose \textbf{ReDI}, a \textbf{reasoning-enhanced} query understanding model that combines both \textbf{decomposition} and \textbf{interpretation}. 
We design a three-stage \ac{LLM}-based pipeline:
\begin{enumerate*}[label=(\roman*)]
\item Decompose a complex query into sub-queries to cover diverse user intents.
\item Augment each sub-query with semantic interpretation to improve alignment between intent and document content.
\item Fuse the retrieval results with a dedicated aggregation strategy to produce the final ranking.
\end{enumerate*}
Moreover, we introduce tailored query prompts for both sparse and dense retrieval, enhancing ReDI’s adaptability across different settings. By explicitly decomposing complex queries and interpreting each sub-intent, ReDI achieves broader coverage and more precise retrieval, delivering strong performance even with simple models such as BM25 and SBERT.

\heading{A new dataset for \acl{QU}}
To support research in this area, we curated a large-scale dataset of complex queries filtered from commercial search logs. Using DeepSeek-R1, we generated high-quality intent annotations to distill a compact student model for production environments, ensuring efficiency and privacy.

Experiments on BEIR \cite{Thakur2021BEIRAH} and BRIGHT \cite{su2025bright} show that ReDI consistently outperforms strong QU baselines in sparse and dense retrieval. Furthermore, our distilled model matches or exceeds its teacher's performance in generating intent-aware queries, validating its scalability and practical utility.

\heading{Main contributions}
We have three main contributions:
\begin{itemize}[leftmargin=*]
    \item We introduce \textbf{ReDI}, a model that enhances complex query decom\-position with complementary interpretations, achieving strong results using lightweight retrieval methods.
    \item We release a large-scale, real-world complex query dataset and distill the \acl{QU} capabilities of DeepSeek-R1 into a lightweight, production-ready model.
    \item Extensive experiments on BEIR and BRIGHT show that ReDI outperforms strong baselines and demonstrates robust generalization across datasets and tasks.
\end{itemize}

\section{Related Work}
\subsection{Traditional query understanding}
Traditional \ac{QU} methods have aimed to mitigate the lexical mismatch problem by enriching queries with additional relevant terms such as synonyms, terms on the same topic, and words with the same root. These approaches typically fall into two main categories based on the sources used: external knowledge-based and \ac{PRF} methods. First, external knowledge-based approaches expand the original query by incorporating semantically related terms from resources like WordNet or Wikipedia \cite{Voorhees94QELR,10.1145/2513204.2513209,10.5555/1625275.1625535}.
 
For example, \citet{Voorhees94QELR} uses WordNet to expand semantically similar terms, and \citet{10.5555/1625275.1625535} employ explicit semantic analysis to embed queries into a Wikipedia‑derived concept space.
Some researchers also employ anchor text \cite{10.1145/1718487.1718493, 10.1145/988672.988763} and user logs \cite{10.1109/TKDE.2003.1209002, Wang2008MiningTA} to extract related terms. 
 
Second, \ac{PRF} approaches use top-ranked pseudo-relevant documents from initial retrieval to derive expansion terms~\cite{ 10.1145/383952.383972,10.1109/WI-IAT.2012.226,10.1108/eb026866}.
%, often through methods like Rocchio feedback \cite{Rocchio1971RelevanceFI} or probabilistic models \cite{10.1108/eb026866}. 
For example, \citet{10.1145/366836.366860} employ \ac{KLD} to assess the differences in term distribution between pseudo-relevant documents and the complete document set.

Despite their effectiveness in specific scenarios, these methods have limitations such as reliance on predefined static semantic resources or susceptibility to semantic drift resulting from the quality of initial retrieval sets \cite{10.1145/2009916.2009942,10.1145/1645953.1646059}.

\subsection{LLM-based query understanding}
Recent advancements in \acp{LLM} have paved the way for novel \ac{QU} approaches that exploit the generative capabilities of these models \cite{Gao2022PreciseZD,Wang2023Query2docQE,Ma2023QueryRF}. Methods such as HyDE \cite{Gao2022PreciseZD} and Query2Doc \cite{Wang2023Query2docQE} use \acp{LLM} to generate hypothetical documents or pseudo-answers, significantly enhancing the semantic richness of queries. 
RRR~\cite{Ma2023QueryRF} uses \acp{LLM} to train a small rewriting model via reinforcement learning, while RAG-STAR~\cite{jiang2024rag} integrates retrieved information to guide a tree-based decomposition process. RQ-RAG~\cite{chan2024rq} enhances models by equipping them with capabilities for explicit rewriting, decomposition, and disambiguation. STEP-BACK~\cite{zheng2024take} performs abstractions to derive high-level concepts and first principles from the original query. \citet{Lei2025ThinkQEQE} iteratively refine expansions using retrieval feedback from the corpus. 
On short queries, LLM-based \ac{QU} methods have shown improved alignment with relevant documents compared to traditional approaches.

\subsection{Reasoning-intensive retrieval}
Complex queries, which often involve multiple entities, broader temporal scopes, and diverse knowledge domains, pose significant challenges to existing \ac{QU} methods. The recently introduced BRIGHT benchmark \cite{su2025bright} offers a structured framework for evaluating \ac{QU} techniques on such queries, which require intensive reasoning to retrieve relevant information.

To tackle these challenges, recent studies have explored using the reasoning capabilities of \acp{LLM} to enhance retrieval effectiveness, broadly falling into two categories.
The first focuses on \textbf{reasoning-augmented ranking}, where ranking models are built upon reasoning-oriented \acp{LLM} \cite{Shao2025ReasonIRTR,liu2025reasonrank,Long2025DIVERAM}. For example, ReasonIR \cite{Shao2025ReasonIRTR} constructs a reasoning-intensive retrieval dataset to train a Llama3.1-based retriever, while ReasonRank \cite{liu2025reasonrank} develops a reasoning-enriched dataset to train Qwen2.5 as a listwise reranker.
The second category uses \acp{LLM} for \textbf{reasoning-based query understanding}, enhancing query representations through generative augmentation \cite{Qin2025TongSearchQRRQ, Long2025DIVERAM, Lei2025ThinkQEQE}. For example, \citet{Long2025DIVERAM} design a feedback-driven query expansion process inspired by \cite{Lei2025ThinkQEQE}. These approaches iteratively refine the query based on the retrieved context.

Our work aligns with the latter line of research but takes a distinct perspective: rather than drilling down into the query’s intent, we focus on \textit{decomposing} the query to uncover and address its multi-dimensional information needs.

\begin{figure*}[!t]
  \centering
  \includegraphics[width=1\textwidth]{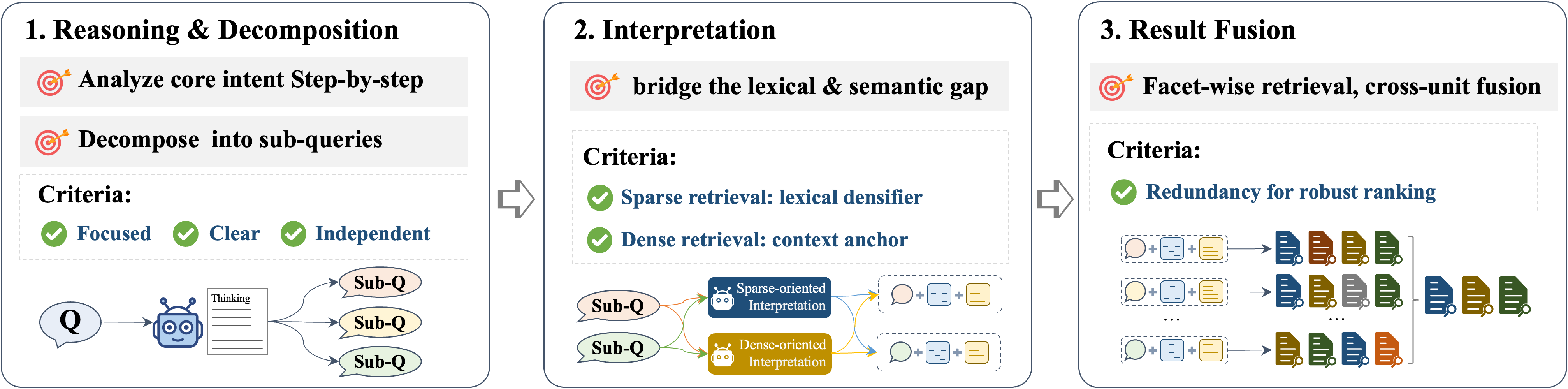}
  \caption{The three-step ReDI workflow.}
  \label{fig:method_main_pic}
\end{figure*}

\section{Methodology}
\label{sec:method}
We propose \textbf{ReDI}, a structured \acl{QU} model that employs \acp{LLM} to systematically process complex queries through three distinct stages: (i)~\textbf{intent reasoning and decomposition}, where the query is analyzed and broken down into focused sub-queries; (ii)~\textbf{sub-query interpretation generation}, where each sub-query is enriched with alternative phrasings and additional contextual information; and (iii)~\textbf{retrieval result fusion}, where each enriched sub-query is independently retrieved, and their results are combined through a fusion strategy into a final ranking. 
Figure~\ref{fig:method_main_pic} illustrates the overall workflow of ReDI. Below, we detail each component.

\subsection{Intent reasoning and query decomposition}
\label{sec:reasoning and decomposition}
Complex queries frequently encompass multiple implicit sub-intents and require multi-hop \ac{IR} from various sourc\-es~\cite{wolfson-etal-2020-break}. Treating these queries as a single retrieval unit often leads to incomplete results~\cite{10.1145/1498759.1498766}. To mitigate this, we first explicitly identify the underlying intent of the original query and decompose it into targeted, independently retrievable sub-queries.

Specifically, given a query $q_i \in \mathcal{Q}$, we first prompt an \ac{LLM} to uncover what the user fundamentally seeks. By \textbf{reasoning} about the core intent, the model identifies whether the query is composed of multiple sub-intents or logical components. We then guide the model to dynamically \textbf{decompose} $q_i$ into a set of clear, concise, and independent sub-queries $S_i = \{s_1$,  $s_2$, \dots, $s_m\}$, each corresponding to a specific aspect of the overall information need. This explicit decomposition ensures thorough coverage of the multi-hop or multi-faceted nature inherent in complex queries, enabling targeted retrieval of documents relevant to each distinct facet. 

% For example, the query ``Should I divest my holdings in Company A before next quarter’s earnings?'', \textbf{ReDI} first identifies the core intent as assessing Company A's investment risk. It then decomposes the query into four focused sub-questions associated with different intents, such as ``A's Production Challenges'', ``Equity Price Volatility'', ``Macroeconomic Factors'', and ``Market Competition''. By handling and retrieving each sub-query individually, the retrieval system efficiently gathers comprehensive documents covering the overall information needs of the original query.

\subsection{Adaptive sub-query interpretation}
\label{sec:interpretation}
Solely relying on decomposition often yields sub-queries that are logically sound but semantically fragmented. These fragmented units suffer from underspecification, failing to provide sufficient signals for effective retrieval. We posit that the nature of this ``information gap'' differs fundamentally across retrieval paradigms. Therefore, an adaptive interpretation strategy is necessary to explicitly bridge these distinct forms of underspecification.

\heading{Lexical underspecification (sparse retrieval)} Sparse retrievers such as BM25 are primarily driven by exact term overlap, making them particularly sensitive to the vocabulary mismatch problem. In this setting, decomposed sub-queries often fail to adequately capture the lexical space of the underlying information need. To address this, our sparse-oriented interpretation functions as a \textit{lexical densifier}, enriching each sub-query with synonyms, morphological variants, and closely related entities. This expansion increases surface-form coverage while preserving the original semantic intent, thereby improving recall under term-based matching.

\heading{Semantic underspecification (dense retrieval)} Dense retrievers encode queries and documents into a shared embedding space, where retrieval quality depends on semantic alignment rather than exact wording. Fragmented or overly concise sub-queries may produce embeddings that poorly reflect the intended conceptual context, leading to misalignment with relevant document clusters. Our dense-oriented interpretation acts as a \textit{contextual anchor}, reformulating the sub-query into fluent, descriptive prose that situates the information need within an appropriate semantic frame (e.g., domain context, causal relations, or task perspective). This anchoring effect guides the query embedding toward regions of the vector space occupied by relevant passages.

\heading{Intent underspecification (reasoning-level interpretation)} Beyond lexical and semantic realization, decomposed sub-queries may also omit the implicit rationale or assumptions underlying the user’s information needs. We therefore prompt the \ac{LLM} to generate a brief intent-grounded interpretation that explicates why the sub-query is being asked and what role it plays in the overall reasoning chain. This reasoning-level context provides an additional retrieval signal, encouraging the retriever to favor passages that align with the deeper intent rather than superficial similarity alone.

\smallskip\noindent%
Taken together, these adaptive interpretations address complementary dimensions of underspecification, enabling robust retrieval across both sparse and dense paradigms.

\subsection{Retrieval result fusion}
% Result Fusion
\label{sec:result_fusionl}
Previous \ac{QU} approaches, such as the reasoning-expansion introduced by BRIGHT \cite{su2025bright}, typically use  \ac{LLM}-generated reasoning as a single long-form expanded query. However, such lengthy queries easily introduce excessive noise, dilute key term importance, and confuse the retrieval model~\cite{10.5555/3666122.3667790}.
To mitigate these issues, we decompose each complex query into multiple sub-queries and enrich them with detailed explanations. This allows each sub-query to focus on a specific aspect of the original query, enabling more precise retrieval with lightweight retrievers.

\heading{Sparse retrieval}
In sparse retrieval, each retrieval unit is independently scored using the BM25 function. Given a sub-query $s_i$ with its interpretation $e_i$ and a candidate document $d$, we construct the query representation by simple concatenation:
\begin{equation}
    \hat{s_i} = s_i \oplus e_i.
\end{equation}
% We adopt simple concatenation without explicit weighting. 
The trade-off between a query and its expansion has been investigated in prior studies \cite{Clinchant2013Aggregating, 10.1145/3627673.3679930}, and we leave its exploration to future work.
%We leave the exploration of adaptive weighting strategies for future work.
\begin{equation}
%\resizebox{0.95\columnwidth}{!}{$
\begin{split}
&\text{Sparse}(\hat{s_i}, d) = \sum_{t \in \hat{s_i} \cap d} \text{IDF}(t) \cdot {}\\
&\mbox{}\hspace*{1.5cm}
\frac{f_d(t) \cdot (k_1 + 1)}{f_d(t) + k_1 \cdot \left(1 - b + b \cdot \frac{|d|}{\text{avgdl}}\right)} \cdot \frac{f_{\hat{s_i}(t)} \cdot (k_3 + 1)}{f_{\hat{s_i}(t)} + k_3},
\end{split}
%$}
\label{eq:sparse}
\end{equation}
where the $\oplus$ operator denotes the concatenation of two texts, $f_d(t)$ and $f_{\hat{s_i}}(t)$ denote the frequency of term $t$ in document $d$ and $\hat{s_i}$, respectively; $|d|$ is the document length, $\text{avgdl}$ is the average document length in the corpus, and $\text{IDF}(t)$ is the inverse document frequency. 
 The hyperparameters $k_1$, $b$, and $k_3$ control document term frequency scaling, length normalization, and query term frequency saturation, respectively.

In particular, we emphasize the role of $k_3$, which controls the impact of query-side term frequency. A smaller $k_3$ amplifies the effect of repeated key terms, making the retriever more sensitive to core lexical cues, whereas a larger $k_3$ reduces saturation, favoring broader coverage across terms. Although many modern BM25 implementations omit this parameter, prior work has noted its potential impact, especially for long and reasoning-intensive queries~\cite{Lin2007Pubmed}. Recent analyses further highlight how different BM25 implementations, particularly whether query-side term frequency is saturated or not, can lead to notable effectiveness differences on BRIGHT~\cite{Ge2025LightingWayBright}. We provide a detailed study of the impact of $k_3$ in the hyperparameter sensitivity experiments reported in Section~\ref{sec:ablation_finetune}.

\heading{Dense retrieval}
For dense retrieval, we follow the prior bi-encoder model DPR~\cite{Karpukhin2020DPR} to encode each sub-query and its corresponding interpretation using a shared dense encoder $f(\cdot)$. A fused query embedding is constructed as a weighted combination of the two, and its similarity to a document embedding is computed via inner product:
\begin{equation}
\mbox{}
\hspace*{-1mm}
\resizebox{0.75\columnwidth}{!}{$
\text{Dense}(s_i, e_i, d) = \left\langle \lambda \cdot f(s_i) + (1 - \lambda) \cdot f(e_i), f(d) \right\rangle
$},
\hspace*{-1mm}\mbox{}
\label{eq:dense}
\end{equation}
where $s_i$ and $e_i$ are the $i$-th sub-query and its interpretation, respectively. The scalar $\lambda \in [0, 1]$ adjusts the relative contribution of the original sub-query semantics and the enriched interpretation. This formulation enables the retrieval model to attend both to the core information need and its contextual elaboration.

\heading{Fusion strategy}
We aggregate the independent scores from all retrieval units to derive the final ranking. Let $\{(s_i, e_i)\}_{i=1}^m$ denote the pairs of sub-queries and interpretations. The final relevance score for document $d$ is computed via summation:
%Once all retrieval units have been independently scored, we aggregate the results to compute the final document score. Let $S = \{s_1, s_2, \dots, s_m\}$ and $E = \{e_1, e_2, \dots, e_m\}$ denote the set of $m$ sub-queries and corresponding interpretations for query $q$. The final relevance score for a document $d$ is computed by summing its scores across all units:
%
\begin{equation}
\text{score}(q, d) = \sum_{s_i \in S, e_i \in E} \text{Retrieval}(s_i, e_i, d),
\end{equation}
where $\text{Retrieval}(s_i, e_i, d)$ corresponds to either $\text{Sparse}(s_i, e_i, d)$ defined in Eq.~\ref{eq:sparse} or $\text{Dense}(s_i, e_i, d)$ defined in Eq.~\ref{eq:dense}. 
We deliberately adopt this \textbf{parameter-free summation strategy} over learning-to-rank approaches. Summation acts as a robust voting mechanism where documents addressing multiple sub-intents naturally accrue higher scores. This approach minimizes the risk of overfitting to specific query structures, ensuring strong zero-shot generalization as demonstrated in Figure~\ref{fig:fusion_method} below. Additional analyses of alternative fusion strategies further corroborate this observation.
%This additive fusion approach prioritizes documents that are relevant to multiple retrieval units, thereby capturing the compositional structure of complex queries and aligning more faithfully with the user’s complete information need.
% We also studied several alternative fusion strategies, which are discussed in Appendix~\ref{sec:supplementary_exp}.

\section{Dataset Collection \& Model Training}
\label{sec:creation}
To be able to train ReDI for accurate understanding, decomposition, and interpretation of complex queries, we construct a dataset of real user queries that naturally embody multifaceted intents. Based on this dataset, we conduct knowledge distillation to develop compact models with enhanced complex \acl{QU} capabilities.

\subsection{Creation of the \textsc{Coin} dataset}
\label{sec:coin_creation}
%With the rise of AI-based search, user queries are shifting from short keyword strings to longer, intent-driven formulations. However, existing datasets mainly focus on simple or synthetic queries, failing to capture real-world complexity. 

\begin{figure}[H]
\centering
\includegraphics[width=\columnwidth]{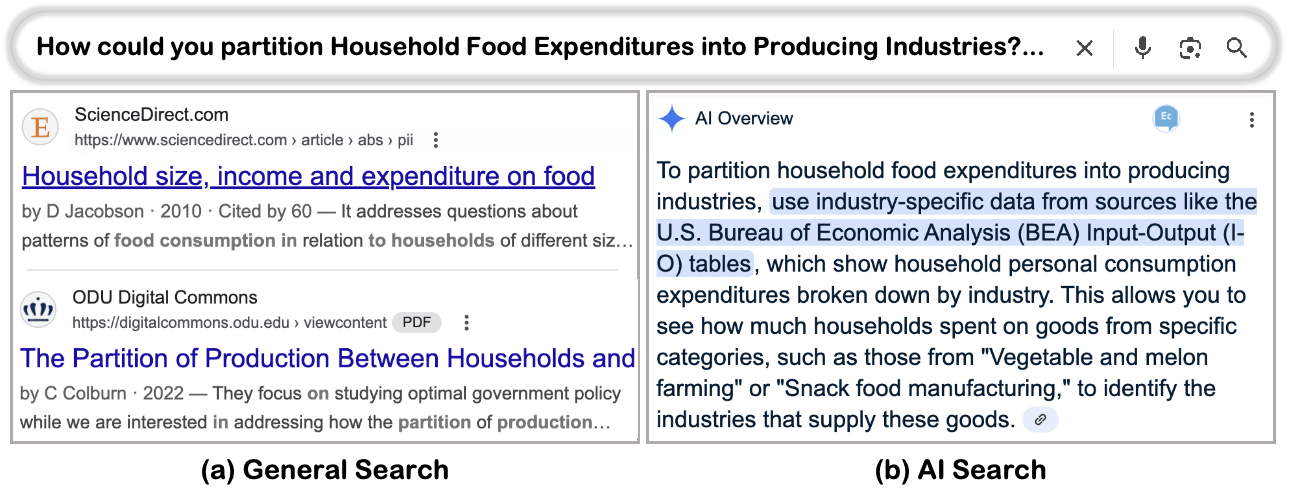}
\caption{General search vs.\ AI search illustration.}
\label{fig:gen_vs_ai_search}
\end{figure}

We propose the \textbf{C}omplex \textbf{O}pen-domain \textbf{IN}tent (\textbf{\textsc{Coin}}) dataset, which targets complex queries from a major search engine.
Drawing from real search logs, the \textsc{Coin} dataset reflects genuine user needs that are \textbf{open-domain} (spanning diverse topics) and \textbf{complex} (involving multiple reasoning steps or aspects). We construct the \textsc{Coin} dataset from two sources: about 100k queries from \textbf{general search}, which emphasize challenging single-turn queries for \textbf{information-locating retrieval}, and about 10k queries from \textbf{AI search}, which emphasize multi-turn queries for \textbf{reasoning-intensive retrieval}. Their differences are illustrated in Figure~\ref{fig:gen_vs_ai_search}.

\begin{figure}[H]
\centering
\includegraphics[width=0.9\columnwidth]{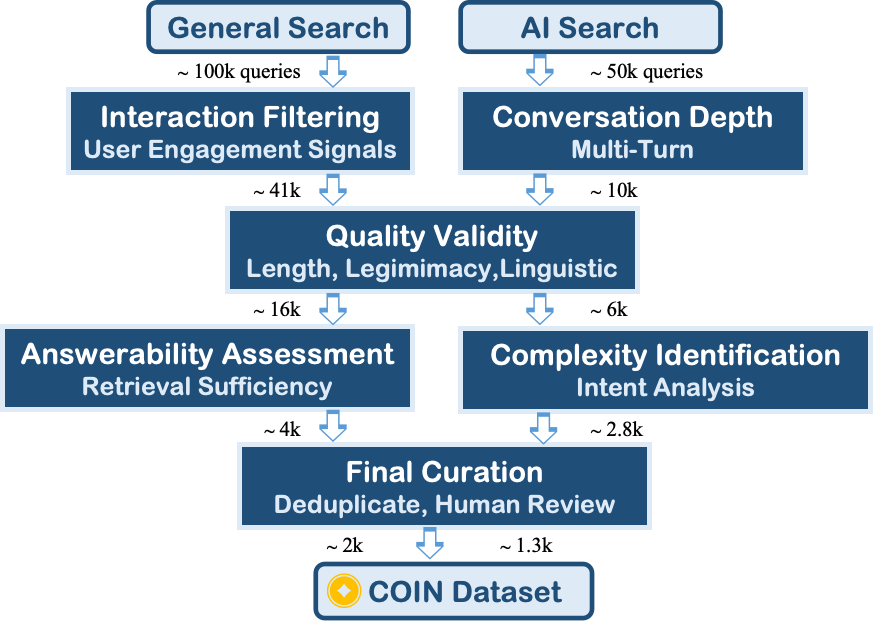}
\caption{The \textsc{Coin} selection workflow.}
\label{fig:final-selection-workflow}
\end{figure}

\noindent%
As shown in Figure~\ref{fig:final-selection-workflow}, the construction of \textsc{Coin} proceeds through four stages. We begin with \textbf{source-specific filtering}, where general search queries are pruned by engagement signals (reducing $\sim$100k to $\sim$41k) and AI search queries are retained only if they exhibit substantive multi-turn context (reducing $\sim$10k to $\sim$10k). Next, a \textbf{quality validation} stage ensures adequate length, linguistic clarity, and legitimacy; DeepSeek-R1 is used to verify fluency and remove trivial, incomplete, or sensitive queries, leaving $\sim$16k and $\sim$6k queries, respectively. This is followed by \textbf{complexity refinement}, where general search queries solvable by top-4 retrievals are discarded ($\sim$4k remain), while AI search queries are filtered by a complex intent classifier to preserve those requiring multi-dimensional reasoning, comparison, or synthesis ($\sim$2.8k remain). Finally, a \textbf{curation} stage merges and de-duplicates both sources, with manual review ensuring diversity and representativeness, yielding 2,056 general search queries and 1,347 AI search queries. In total, the \textsc{Coin} dataset contains 3,403 unique complex queries.

\heading{\textsc{Coin} dataset validation}
To verify that the selected \textsc{Coin} queries indeed require decomposition, we conduct a comparative answering experiment on \emph{retained} (complex) versus \emph{excluded} (simple) queries. For each query, we retrieve the top-4 documents via a standard search API and prompt DeepSeek-R1 to generate an answer by synthesizing information from those documents. We then evaluate each answer along four key dimensions of quality: \textbf{Accuracy (Acc.)} (correctness of the information), \textbf{Completeness (Compl.)} (coverage of all aspects of the query), \textbf{Coherence (Coh.)} (logical consistency and fluency), and \textbf{Conciseness (Conc.)} (absence of unnecessary or off-topic content). Each dimension was rated on a 1–5 scale by DeepSeek-R1 judger, and we averaged these ratings to obtain an overall QA score for the query.

As shown in Table~\ref{tab:dataset-validation}, excluded queries achieved high QA scores (3.65/5), indicating that a single round of retrieval and \ac{LLM} answering often sufficed for these queries. In contrast, our \textsc{Coin} dataset retained queries scored much lower on average (1.95/5), with particularly poor performance on completeness (1.9/5). This result confirms that \textsc{Coin}’s queries inherently demand multi-faceted reasoning and are ill-served by straightforward retrieval, underscoring the importance of an intent-decomposition approach.

\begin{table}[H]
  \centering
  \caption{DeepSeek‐R1 average ratings (excluded vs. retained).}
  \label{tab:dataset-validation}
  %\scriptsize               % smaller font
  \setlength{\tabcolsep}{8pt}
  \begin{tabular}{@{}lccccc@{}}
    \toprule
    Type           & Acc. & Compl. & Coh. & Conc. & Avg. \\
    \midrule
    Excluded queries & 3.8    & 3.6          & 3.7       & 3.5         & 3.65 \\
    Retained queries & 2.1    & 1.9          & 2.0       & 1.8         & 1.95 \\
    \bottomrule
  \end{tabular}
\end{table}

\subsection{Efficient model fine-tuning}
\label{sec:model-fine-tuning}
To enable structured intent understanding, we fine-tune models on the \textsc{Coin} dataset for three sub-tasks: query decomposition, sparse-oriented interpretation, and dense-oriented interpretation. As the dataset lacks ground-truth labels, we employ DeepSeek-R1 to generate high-quality annotations and explore two paradigms:

\subsubsection{Two-stage fine-tuning.}
We train a decomposition model and two interpretation models separately:
\begin{itemize}[leftmargin=*]
\item \textbf{Decomposition.} Given a raw query $q$, the model generates a set of sub-queries $\{{s_i,\dots,s_m}\}$, each targeting one facet of the information need.
\item \textbf{Interpretation.} For each $s_i$, two models produce corresponding interpretations $e_i$: (a) Sparse-oriented: focus on lexical richness (synonyms, derivations, domain-specific terms). (b) Dense-oriented: emphasize semantic clarity and paraphrasing.
% \end{itemize}
\end{itemize}
The objective minimizes sequence generation loss:
\begin{equation}
\mathcal{L}_{\text{two-stage}}=\mathbb{E}{q \sim \mathcal{Q}}\left[\sum_{i=1}^N \big(\log P(s_i \mid q)+\log P(e_i \mid s_i)\big)\right].
\end{equation}

\subsubsection{Joint fine-tuning.}
Alternatively, we jointly fine-tune a single model to perform decomposition and interpretation generation in one pass. Given a query $q$, the model outputs interleaved sub-queries and their corresponding interpretations in parallel:
\begin{equation}
q \rightarrow {(s_1,e_1),\dots,(s_m,e_m)}.
\end{equation}
We optimize a joint loss:
\begin{equation}
\mathcal{L}_{\text{joint}}=\alpha\mathcal{L}_{\text{decomp}}+(1-\alpha)\mathcal{L}_{\text{interp}},
\end{equation}
where $\alpha$ controls the balance between the learning of decomposition and contextual relevance.

%encouraging the model to learn both decomposition and contextual relevance.

All fine-tuned models are referred to as \textbf{ReDI}. As shown in Section~\ref{sec:analysis}, both variants achieve strong performance on BRIGHT, rivaling or surpassing larger baselines.

\section{Experiments}
\label{sec:experiment}
\subsection{Experimental setup}
\subsubsection{Datasets}
We evaluate our method on two prominent retrieval benchmarks: \textbf{BRIGHT} and \textbf{BEIR}, covering a wide range of real-world query scenarios.
BRIGHT~\cite{su2025bright} is a reasoning-intensive benchmark with 1,384 real-world queries across \textit{StackExchange}, \textit{Coding}, and \textit{Theorem-based}, including a long-document subset of StackExchange tasks where queries are matched against full-length web pages with higher token counts and noise.
BEIR~\cite{Thakur2021BEIRAH} is a heterogeneous IR benchmark of 18 datasets; following prior work Rank1~\cite{weller2025rank1}, we evaluate on a subset of 9 datasets with fewer than 2,000 queries: (\textit{ArguAna}, \textit{Climate-FEVER}, \textit{DBPedia}, \textit{FiQA-2018}, \textit{NFCorpus}, \textit{SciDocs}, \textit{SciFact}, \textit{Webis-Touche2020}, \textit{TREC-COVID}).

\subsubsection{Metrics}
Following BRIGHT and Rank1, we adopt nDCG@10 as the primary evaluation metric. Specifically, for the long-document subset of BRIGHT, we follow BRIGHT and report Recall@1. We performed significant tests using the paired t-test. Differences are considered statistically significant when the $p$-value is lower than 0.05.

\subsubsection{Baselines}
\label{sec:baseline}
We use the reasoning expansion variants released in the official BRIGHT dataset repository,\footnote{\url{https://huggingface.co/datasets/xlangai/BRIGHT}} generated by Claude-3-opus and GPT-4, as our baselines.
Moreover, we generate reasoning expansions with DeepSeek-R1, similarly to BRIGHT.
In addition, we incorporate several recently proposed state-of-the-art methods, i.e., TongSearch-QR~\cite{Qin2025TongSearchQRRQ}, ThinkQE~\cite{Lei2025ThinkQEQE}, and DIVER-QExpand~\cite{Long2025DIVERAM} as baselines, and reproduce their results.

\subsubsection{Training details.}
We fine-tune Qwen3-8B on the \textsc{Coin} dataset described in Section~\ref{sec:coin_creation}, using a \textit{lr} of $1\times10^{-4}$ with 10 \% linear warm-up and cosine decay. %All experiments are conducted on a single NVIDIA A100 GPU. 

\subsubsection{Evaluation procedure.}
We evaluate \ac{QU} methods under both sparse and dense paradigms, using two representative lightweight retrievers.
%\textbf{ReDI} follows the unit-level strategy in Section~\ref{sec:result_fusion}. 
For Sparse Retrieval, we use Gensim’s LuceneBM25Model\footnote{\url{https://pypi.org/project/gensim/}} with Pyserini\footnote{\url{https://pypi.org/project/pyserini/}} analyzers. Baselines use reasoning-expanded queries with the BRIGHT BM25 setting ($k_1=0.9$, $b=0.4$). ReDI modifies this setup by adjusting $k_3$, setting it to $0.4$ for short documents and $5$ for long ones.
%($k_1{=}0.9, b{=}0.4, k_3{=}0.4$; $k_3{=}5$ for long-docs).
%, retrieving the top-1k documents per unit and summing scores across units.
For Dense Retrieval, we use SBERT,\footnote{\url{https://huggingface.co/sentence-transformers/all-mpnet-base-v2}} a relatively lightweight bi-encoder model that encodes queries and documents into 768-dimensional embeddings.
ReDI embeds each sub-query and interpretation, fuses them by weighted average ($\lambda=0.5$ for BRIGHT, $0.4$ for long documents). For model learning, we set $\alpha=0.5$ to balance the contribution of both decomposition and interpretation.

All experiments are zero-shot: ReDI is only trained on \textsc{Coin} without exposure to BRIGHT or BEIR, ensuring  fair evaluation.

% \subsubsection{Statistical significance test}
% We conduct paired two-sided t-tests over per-query scores between ReDI and each baseline under the same retrieval setting. Statistically significant improvements (p-value $\leq$ 0.05) are marked with $*$ in the result tables.

\begin{table*}[!t]
 \centering
 \caption{nDCG@10 on BRIGHT. Best scores are in bold, second‐best are underlined. Avg.All and Avg.SE denote average performance across all and StackExchange tasks respectively. Significant improvement or degradation with respect to ReDI is indicated (+/-) (p-value $\leq$ 0.05).
}
 \label{tab:main_result_1}

 \setlength{\tabcolsep}{4pt}
 \begin{tabular}{
  l r r r
  r r r r r r r
  r r
  r r r}
  \toprule    
   \multirow{2}{*}{\numb{\ac{QU} Model}}
   & \multirow{2}{*}{\numb{Params}}
   & \multirow{2}{*}{\numb{Avg.All}}
   & \multirow{2}{*}{\numb{Avg.SE}}
   & \multicolumn{7}{c}{\textit{StackExchange}}
   & \multicolumn{2}{c}{\textit{Coding}}
   & \multicolumn{3}{c}{\textit{Theorem‐based}} \\
  \cmidrule(lr){5-11} \cmidrule(lr){12-13} \cmidrule(lr){14-16}
   &
   &
   &
   & Bio. & Earth. & Econ. & Psy. & Rob. & Stack. & Sus.
   & Leet. & Pony
   & AoPS & TheoQ. & TheoT. \\
  \midrule
   %\rowcolor{gray!20} 
   \multicolumn{16}{c}{\emph{Using BM25 Retriever}}\\
   \midrule
   - & - 
   & \numm{14.5} & \numm{17.2} 
   & \numm{18.9} & \numm{27.2} & \numm{14.9} & \numm{12.5} & \numm{13.6} & \numm{18.4} & \numm{15.0} 
   & \num{24.4} & \num{7.9} 
   & \num{6.2} & \numm{10.4} & \numm{4.9} \\
   \midrule
    Claude-3-opus & -
    & \numm{26.8} & \numm{34.4}
    & \num{54.2} & \num{52.1} & \numm{23.5} & \num{38.4} & \numu{22.5} & \numm{24.1}& \num{26.0} 
    & \numm{20.0} & \numup{19.6} 
    & \num{4.1} & \numm{19.0} & \numm{18.1} \\

    GPT4 & -
    & \numm{27.0} & \numm{34.8}
    & \num{53.6} & \numu{54.1} & \num{24.3} & \numu{38.7} & \numm{18.9} & \numm{27.7} & \num{26.3} 
    & \numm{19.3} & \nump{17.6} 
    & \num{3.9} & \numm{19.2} & \numm{20.8} \\

    DeepSeek-R1 & 671B
    & \numum{29.2} & \numu{37.0}
    & \numup{57.2} & \numb{58.1} & \num{24.0} & \num{38.1} & \num{22.1} & \numum{29.6} & \numb{29.6} 
    & \num{22.2} & \nump{12.4} 
    & \num{6.8} & \numum{26.3} & \numu{23.4} \\

   \midrule
   
    TongSearch-QR & 7B
    & \numm{27.9} & \numm{34.5}
    & \numbp{57.9} & \num{50.9} & \numm{21.9} & \numm{37.0} & \num{21.3} & \numm{27.0} & \num{25.6}
    & \num{23.7} & \nump{14.4}
    & \numu{7.0} & \numm{26.1} & \numm{22.0} \\

    ThinkQE & 14B 
    & - & \numm{34.5} 
    & 47.2 & 49.5 & \numb{29.0} & \numm{37.8} & \numm{20.9} & \numm{28.2} & 28.8 & - & - 
    & - & - & - \\
    
    DIVER-QExpand & 14B
    & \numm{27.0} & \numm{33.5}
    & \num{50.9} & \num{50.3} & \numm{22.4} & \numm{35.3} & \num{22.2} & \numm{27.3} & \num{26.4}
    & \numb{26.7} & \numbp{20.6}
    & \numb{8.5} & \numm{19.2} & \numm{14.2} \\

    \textbf{ReDI} (ours) & 8B 
    & \numb{30.8} & \numb{38.3} 
    & \num{49.0} & \num{53.5} & \numu{28.7} & \numb{43.4} & \numb{27.5} & \numb{36.3} & \numu{29.4} 
    & \numu{25.3} & \num{9.3} 
    & \num{6.0} & \numb{31.5} & \numb{30.0} \\
    
  \midrule
  %\rowcolor{gray!20} 
  \multicolumn{16}{c}{\emph{Using SBERT Retriever}}\\
  \midrule
    - & - 
    & \numm{14.9} & \numm{15.6} 
    & \numm{15.1} & \numm{20.4} & \numm{16.6} & \numm{22.7} & \numm{8.2} & \numm{11.0} & \numm{15.3} 
    & \numb{26.4} & \numm{7.0} 
    & \num{5.3} & \numm{20.0} & \numm{10.8} \\
  \midrule
    Claude-3-opus & - 
    & \numm{16.4} & \numm{18.0} 
    & \numm{18.6} & \numm{24.8} & \num{18.6} & \num{24.9} & \numu{11.4} & \numum{12.9} & \numm{14.7} 
    & \num{23.0} & \numm{5.8} 
    & \numm{3.1} & \numm{20.1} & \numm{19.0} \\
    
    GPT4 & - 
    & \numm{17.7} & \numm{18.2} 
    & \numm{18.5}  & \numm{26.3}  & \num{17.5}  & \numu{27.2} & \numm{8.8}  & \numm{11.8}  & \numm{17.5}  
    & \num{24.3}  & \numum{10.3} 
    & \num{5.0} & \numm{22.3}  & \num{23.5}  \\
    
    DeepSeek-R1 & 671B 
    & \numm{18.9} & \numm{19.9} 
    & \num{20.8} & \numu{31.0} & \numu{20.2} & \num{26.0} & \num{10.3} & \numm{12.4} & \numum{18.6} 
    & \numm{22.6} & \numm{4.5} 
    & \numb{8.4} & \numm{27.9} & \num{23.8} \\

  \midrule
  
    TongSearch-QR & 7B  
    & \numm{18.5} & \numm{18.7} 
    & \numm{20.5} & \numm{25.5} & \num{18.4} & \num{25.5} & \num{11.2} & \numm{11.6} & \numm{18.4} 
    & \num{23.4} & \numm{9.5} 
    & \num{4.7} & \numum{28.0} & \numu{25.2} \\
    
    DIVER-QExpand & 14B 
    & \numum{19.2} & \numum{20.3} 
    & \numbp{28.3} & \numm{27.6} & \num{20.0} & \num{26.6} & \num{10.5} & \numm{11.7} & \num{17.0} 
    & \numu{26.1} & \numm{3.4} 
    & \numu{6.4} & \numm{26.9} & \numu{25.2} \\
    
    \textbf{ReDI} (ours)  & 8B 
    & \numb{22.8} & \numb{23.7} 
    & \numu{25.0} & \numb{32.3} & \numb{20.8} & \numb{28.0} & \numb{13.8} & \numb{20.2} & \numb{25.6} 
    & \num{25.2} & \numb{17.1} 
    & \num{6.2} & \numb{33.2} & \num{25.8} \\  
  \bottomrule
 \end{tabular}
 \medskip
\end{table*}

\subsection{Main results}
Table~\ref{tab:main_result_1} reports the retrieval performance of different \ac{QU} methods on BRIGHT. Key observations include the following.

\heading{Sparse retrieval} ReDI consistently improves the performance on BM25, achieving the highest average nDCG@10 of $30.8\%$. Compared to single long-form expansion methods (Claude-3-opus, GPT-4, DeepSeek-R1), ReDI’s structured decomposition and interpretation lead to more targeted retrieval across StackExchange domains. This may be because long-form expansions often introduce redundant reasoning or topic drift. Among the baselines, feedback-based methods such as \textit{ThinkQE} and \textit{DIVER-QExpand} reach competitive results by iteratively refining queries. Notably, ReDI surpasses both despite not using any retrieval feedback, highlighting the crucial role of explicit decomposition for intent understanding. Nevertheless, ReDI struggles slightly on \textit{Pony} and \textit{AoPS}, where feedback-based methods excel. This highlights the opportunity for future work to explore the synergy of our decomposition with iterative refinement. 

\heading{Dense retrieval} With the SBERT retriever, which employs a 768-dimensional bi-encoder, single-query expansions yield limited improvements compared with the sparse setting. This may be due to the relatively shallow nature of the dense embedding space, which limits its ability to capture nuanced reasoning semantics. Despite this, ReDI achieves the best average nDCG@10 of 22.8\%, with substantial gains on tasks such as \textit{Biology} and \textit{StackOverflow}. These findings indicate that structured decomposition and contextual interpretation can effectively compensate for the representational limits of lightweight dense retrievers.
    For more advanced models (e.g., ReasonIR), Section~\ref{sec:ablation_trans} provides a detailed analysis of transferability and model interaction effects.

Overall, ReDI achieves superior performance across both sparse and dense retrieval settings with two lightweight retrievers. These results confirm the effectiveness of decomposition and interpretation for complex query understanding, particularly in domains that require abstract reasoning.
In terms of efficiency, ReDI also offers a strong trade-off: as shown in Table~\ref{tab:efficiency} ( Section~\ref{sec:reasonir}), its \ac{QU} stage (decompose + interpret) is slower than generating single long-form expansions with models of the same size, but substantially faster than using larger models like DeepSeek-R1, while achieving comparable or even better retrieval performance.

\section{Detailed Analyses}
\label{sec:analysis}
To assess the effectiveness of ReDI and understand the rationale behind its design choices, we address the following four research questions:
\heading{RQ1} How do the core components, especially reasoning, decomposition, and interpretation, individually contribute to ReDI's overall retrieval performance? (Section \ref{sec:ablation_modules})
\heading{RQ2} How do different training and retrieval configurations impact ReDI's performance and practical utility? (Section \ref{sec:ablation_finetune})
\heading{RQ3} To what extent does ReDI generalize to different retrieval scenarios, such as long-document contexts and out-of-domain datasets? (Section \ref{sec:ablation_trans})
\heading{RQ4} How does ReDI behave when paired with retrievers that posses intrinsic reasoning capabilities, and does it offer complementary benefits? (Section \ref{sec:reasonir})
%To better understand the effectiveness of our ReDI model, we conduct a comprehensive analysis to dissect the contributions of its core components and design choices. Our analysis consists of four parts: 
%(i)~\textbf{component analysis}: evaluate the individual contributions of reasoning, decomposition, and interpretation to ReDI; 
%% and includes a brief efficiency profile of each stage; 
%(ii)~\textbf{strategy optimization}: examine the impact of different training paradigms and retrieval configurations on performance and practical utility; (iii)~\textbf{transferability evaluation}: assess ReDI's generalization ability on long documents, out-of-domain retrieval; and 
%(iv)~\textbf{Interaction with reasoning-oriented retrievers}: analyzes how ReDI behaves when paired with retrievers that possess intrinsic reasoning capabilities.
%how different training paradigms and retrieval configurations affect model performance and practical utility; and (iii)~\textbf{transferability evaluation}, which assesses the generalization ability of \textbf{ReDI} on long documents, dense retrieval with ReasonIR-8B, and out-of-domain retrieval. 

\smallskip\noindent%
We provide all detailed results in the released code base.\footnote{\url{https://github.com/youngbeauty250/ReDI}}

\begin{table*}[!t]
 \centering
 \caption{Comparison of expansion, decomposition, and decomposition with interpretation on BRIGHT (nDCG@10).}
 \label{tab:decomp_vs_decomp_desc}
  % \small
    \renewcommand{\arraystretch}{0.9}

 \setlength{\tabcolsep}{3.2pt}
 \begin{tabular}{%
  l % Query Rewrite Model
  l % Methods
  r % Avg.All (formerly Avg.)
  r % Avg.ES (formerly StackExchange/Avg.)
  r r r r r r r
  r r
  r r r}
  \toprule    
   \multirow{2}{*}{\textbf{\ac{QU} Model}}
   & \multirow{2}{*}{\textbf{Method}}
   & \multirow{2}{*}{\textbf{Avg.All}}
   & \multirow{2}{*}{\textbf{Avg.SE}}
   & \multicolumn{7}{c}{\textit{StackExchange}}
   & \multicolumn{2}{c}{\textit{Coding}}
   & \multicolumn{3}{c}{\textit{Theorem‐based}} \\
  \cmidrule(lr){5-11} \cmidrule(lr){12-13} \cmidrule(lr){14-16}
   &
   &
   &
   & Bio. & Earth. & Econ. & Psy. & Rob. & Stack. & Sus.
   & Leet. & Pony
   & AoPS & TheoQ. & TheoT. \\
  \midrule
  %\rowcolor{gray!20} 
  \multicolumn{16}{c}{\emph{Using BM25 Retriever}}\\
  \midrule
    \multirow{3}{*}{\textbf{ReDI}}
      & Expansion            & \num{22.6} & 28.8 & 47.0 & \num{47.7} & \num{19.1} & \num{30.4} & \num{15.0} & \num{22.3} & \num{20.0} & \num{18.6} & 7.7 & \num{3.7} & \num{23.1} & \num{16.2} \\
      & Decomp.              & \num{20.7} & 25.2 & \num{26.9} & \num{35.4} & \num{20.2} & \num{26.8} & \num{19.2} & \num{27.6} & \num{20.0} & \num{20.8} & \num{3.7} & \num{3.9} & \num{22.7} & \num{21.5} \\
      & Decomp.+Interp.      & \numb{30.8} & \numb{38.3} & \numb{49.0} & \numb{53.5} & \numb{28.7} & \numb{43.4} & \numb{27.5} & \numb{36.3} & \numb{29.4} & \numb{25.3} & \numb{9.3} & \numb{6.0} & \numb{31.5} & \numb{30.0} \\
      \midrule
      \multirow{3}{*}{DeepSeek-R1}
      & Expansion              & \num{29.2} & \num{37.0} & \numb{57.2} & \numb{58.1} & \num{24.0} & \num{38.1} & \num{22.1} & \num{29.6} & \numb{29.6} & \numb{22.2} & 12.4 & \numb{6.8} & \num{26.3} & \num{23.4} \\
      & Decomp.              & \num{21.3} & \num{26.1} & \num{33.9} & \num{35.6} & \num{22.7} & \num{30.6} & \num{17.2} & \num{23.9} & \num{19.0} & \num{15.6} & \num{5.8} & \num{3.8} & \num{25.0} & \num{22.8} \\
      & Decomp.+Interp.      & \numb{31.9} & \numb{39.9} & 56.6 & 56.4 & \numb{31.7} & \numb{41.8} & \numb{26.3} & \numb{36.8} & 29.4 & 21.2 & \numb{13.5} & 6.3 & \numb{30.6} & \numb{32.0} \\

  \midrule
  %\rowcolor{gray!20} 
  \multicolumn{16}{c}{\emph{Using SBERT Retriever}}\\
  \midrule
      \multirow{3}{*}{\textbf{ReDI}}
      & Expansion            & \num{18.4} & \num{19.2} & \num{20.0} & \num{28.4} & \num{18.4} & 26.2 & 11.2 & \num{14.2} & 16.0 & 24.4 & \num{6.6} & 4.7 & \num{25.5} & 25.2 \\
      & Decomp.              & 20.2 & \num{20.1} & \num{22.4} & \num{25.1} & \num{17.4} & \num{24.3} & 11.6 & 17.8 & \num{22.2} & 24.4 & \numb{17.9} & 3.5 & 31.8 & 23.9 \\
      & Decomp.+Interp.      & \numb{22.8} & \numb{23.7} & \numb{25.0} & \numb{32.3} & \numb{20.8} & \numb{28.0} & \numb{13.8} & \numb{20.2} & \numb{25.6} & \numb{25.2} & 17.1 & \numb{6.2} & \numb{33.2} & \numb{25.8} \\
      \midrule
      \multirow{3}{*}{DeepSeek-R1}
      & Expansion              & \num{18.9} & \num{19.9} & \num{20.8} & \numb{31.0} & 20.2 & 26.0 & 10.3 & \num{12.4} & \num{18.6} & \numb{22.6} & \num{4.5} & \numb{8.4} & \num{27.9} & \num{23.8} \\
      & Decomp.              & 21.0 & \num{20.9} & \num{22.3} & 30.3 & 20.5 & 25.3 & 11.0 & 16.8 & \num{20.2} & 20.9 & \numb{22.0} & 5.9 & \num{30.7} & \num{25.6} \\
      & Decomp.+Interp.      & \numb{22.1} & \numb{22.7} & \numb{25.1} & \numb{31.0} & \numb{21.9} & \numb{26.6} & \numb{12.3} & \numb{18.7} & \numb{23.0} & 18.9 & 18.2 & 4.4 & \numb{35.2} & \numb{29.2} \\
    \bottomrule

 \end{tabular}
 \medskip
\end{table*}

\subsection{Effectiveness of core components}
\label{sec:ablation_modules}
We begin by analyzing how each component, reasoning, decomposition, and interpretation, contributes to performance gains and whether their combination yields additive benefits. We also profile the runtime of each stage to quantify computational cost.

% Ablation Part 1.1 Model Size
\heading{Contribution of the reasoning}
\label{sec:model_reasoning}
We compare Qwen3 models of varying sizes (0.6B/4B/8B) in both \textit{without-thinking} (direct answer) and \textit{with-thinking} (reasoning-augmented) modes to assess how the model’s intrinsic reasoning capacity affects downstream processing. As shown in Figure~\ref {fig:model_size}, both increased model size and explicit reasoning traces lead to consistent gains in retrieval performance on BRIGHT. It is noteworthy that the advantage of incorporating reasoning scales with the base model's size, suggesting that a more powerful base model maximizes the utility of decomposition and interpretation.
%indicating that stronger underlying reasoning capacity amplifies the downstream utility of decomposition and interpretation. 
These results underscore that effective retrieval for complex queries hinges on models that can reason before retrieving.

\begin{figure}[h]
  \centering
  \captionsetup[subfigure]{skip=-2pt}
  \begin{tikzpicture}
    \begin{axis}[
      hide axis,
      xmin=0, xmax=1, ymin=0, ymax=1,
      legend style={
        at={(0.5,1)},    % 这里的 (0.5,1) 表示在 tikzpicture 的顶部中间
        anchor=south,
        legend columns=3,
        draw=none,
        font=\scriptsize
      }
    ]
      \addlegendimage{area legend, fill=MidnightBlue, postaction={pattern=north east lines}}
      \addlegendentry{w/o thinking}
      \addlegendimage{area legend, fill=SeaGreen!60, postaction={pattern=vertical lines}}
      \addlegendentry{w/ thinking}
      \addlegendimage{Red!90,dashed,thick}
      \addlegendentry{baseline}
    \end{axis}
  \end{tikzpicture}
  \vspace{-0.25em} % 根据需要微调 legend 和柱状图之间的距离

  \makebox[1\columnwidth][c]{%
  %——— 左子图：BM25 ———
  \begin{subfigure}[t]{0.6\columnwidth}
    % \hspace*{-10em}
    \centering
      % --- 共用图例 ---
    \begin{tikzpicture}
      \begin{axis}[
 ybar,
 axis on top,
        height=0.9\linewidth,
        width=\linewidth,
        bar width=0.35cm,
        ytick={12,16,20,24},
        ymajorgrids, 
        tick align=inside,
        major grid style={draw=white},
        minor y tick num={0},
        %tickwidth=0pt,
        enlarge y limits={value=.001,upper},
        ymin=12, ymax=25,
        %axis x line*=bottom,
        %axis y line*=left,
        axis lines=left,
        %y axis line style={opacity=0},
        %tickwidth=0pt,
        clip=false,
        enlarge x limits=0.25,
        % legend style={
        %     at={(0.5,1.1)},
        %     font=\small,
        %     anchor=north,
        %     draw=none,
        %     legend columns=-1,
        %     /tikz/every even column/.append style={column sep=0.01cm}
        % },
        symbolic x coords={0.6B, 4B, 8B},
       xtick=data,
       nodes near coords={
        \pgfmathprintnumber[fixed zerofill,precision=1]{\pgfplotspointmeta}
       },
       every node near coord/.append style={anchor=south, font=\fontsize{5pt}{4pt}\selectfont},
       yticklabel style={/pgf/number format/fixed, xshift=1.5pt, /pgf/number format/precision=2, font=\fontsize{6pt}{4pt}\selectfont},
       xticklabel style={rotate=0, anchor=center, yshift=-4pt,font=\fontsize{7pt}{4pt}\selectfont},
    ]
    \addplot [draw=none, fill=MidnightBlue, postaction={pattern=north east lines}] coordinates {
      (0.6B, 15.3)
      (4B, 19.0)
      (8B, 20.7)
    };
    \addplot [draw=none, fill=SeaGreen!60, postaction={pattern=vertical lines}] coordinates {
      (0.6B, 15.6)
      (4B, 20.1)
      (8B, 22.8)
    };
    \draw[Red!90,dashed,thick]
    (axis description cs:0,0.1923) -- (axis description cs:1,0.1923);
    % 在最右侧稍微偏右一点画出标签
    \node[Red!90,anchor=east,font=\fontsize{6.5pt}{5pt}\selectfont]
    at (axis description cs:0,0.1923) {14.5};
    
    \end{axis}
    \end{tikzpicture}
    \caption{Sparse}
    \label{fig:bm25}
  \end{subfigure}%
  \hspace{-2em}
  %——— 右子图：SBERT ———
  \begin{subfigure}[t]{0.6\columnwidth}
    % \hspace*{10em}
    % \centering
    \begin{tikzpicture}
      \begin{axis}[
 ybar,
 axis on top,
        height=0.9\linewidth,
        width=\linewidth,
        bar width=0.35cm,
        ytick={6,10,14,18},
        ymajorgrids, 
        tick align=inside,
        major grid style={draw=white},
        minor y tick num={0},
        %tickwidth=0pt,
        enlarge y limits={value=.001,upper},
        ymin=6, ymax=19,
        %axis x line*=bottom,
        %axis y line*=left,
        axis lines=left,
        %y axis line style={opacity=0},
        %tickwidth=0pt,
        clip=false,
        enlarge x limits=0.25,
        % legend style={
        %     at={(0.5,1.1)},
        %     font=\small,
        %     anchor=north,
        %     draw=none,
        %     legend columns=-1,
        %     /tikz/every even column/.append style={column sep=0.01cm}
        % },
        % ylabel={Average NDCG@10},
        % ylabel style={
        %     anchor=south,
        %     at={(ticklabel* cs:0.5)},
        %     yshift=10pt
        % },
        symbolic x coords={0.6B, 4B, 8B},
       xtick=data,
       nodes near coords={
        \pgfmathprintnumber[fixed zerofill,precision=1]{\pgfplotspointmeta}
       },
       every node near coord/.append style={anchor=south, font=\fontsize{5pt}{4pt}\selectfont, yshift=1.5pt},
       yticklabel style={/pgf/number format/fixed, xshift=1.5pt, /pgf/number format/precision=2, font=\fontsize{6pt}{4pt}\selectfont},
       xticklabel style={rotate=0, anchor=center, yshift=-4pt, font=\fontsize{7pt}{4pt}\selectfont},
    ]
    \addplot [draw=none, fill=MidnightBlue, postaction={pattern=north east lines}] coordinates {
      (0.6B, 10.7)
      (4B, 13.7)
      (8B, 15.3)
    };
    \addplot [draw=none, fill=SeaGreen!60, postaction={pattern=vertical lines}] coordinates {
      (0.6B, 10.3)
      (4B, 14.3)
      (8B, 16.8)
    };
    \draw[Red!90,dashed,thick]
    (axis description cs:0,0.6846) -- (axis description cs:1.,0.6846);
    % 在最右侧稍微偏右一点画出标签
    \node[Red!90,anchor=east,font=\fontsize{6.5pt}{5pt}\selectfont]
    at (axis description cs:0,0.6846) {14.9};
      \end{axis}
    \end{tikzpicture}
    \caption{Dense}
    \label{fig:sbert}
  \end{subfigure}
  }%end makebox

  \caption{nDCG@10 on BRIGHT with Qwen3 across different model sizes and reasoning modes.}
  \label{fig:model_size}
\end{figure}

% Ablation Part 1.2 Expand vs Decomp vs Decomp+Desc
\heading{Contribution of interpretation on decomposition}
We compare three strategies: (a) single long-form expansion (``Expansion''), (b) sub-query decomposition only (``Decomp.''), and (c) decomposition with interpretation (``Decomp.+Interp.''), to assess the added value of enriching each sub-query with contextual interpretation. As shown in Table~\ref{tab:decomp_vs_decomp_desc}, across both retrieval paradigms and generation models, the decomposition plus interpretation approach achieves the highest nDCG@10 on nearly all tasks and in overall averages. The results highlight that decomposition alone is insufficient: adding interpretation significantly improves retrieval by providing semantic grounding, reducing lexical mismatch, and enabling more complete coverage of complex, multifaceted queries.

% Ablation Part 1.3 Unit Len
\heading{Flexible vs.\ fixed decomposition}
We compare fixed and flexible decomposition performance. As shown in Figure~\ref{fig:unit-len}, ReDI with flexible decomposition consistently outperforms all fixed settings under different retrieval models. 
These results highlight the benefit of tailoring decomposition granularity to query complexity, allowing more retrieval units for complex queries and fewer for simpler ones, which improves overall retrieval performance.
% -- allocating more retrieval units to information-dense queries and fewer to simpler ones -- thereby improving retrieval effectiveness across the board. 
Although ReDI offers a flexible way to determine the number of decompositions dynamically, it is worth investigating more targeted methods for optimizing the decomposition for each specific query in the future.

% % Ablation Part 1.4 Efficiency
% \heading{Efficiency Analysis}
% We evaluate the computational efficiency of each stage in the ReDI pipeline on an A100 GPU.
% Decomposition takes an average of \textbf{1.76s} per query, while interpretation generation requires \textbf{7.866s} for sparse-oriented and \textbf{5.983s} for dense-oriented models.
% Retrieval and fusion are lightweight -- \textbf{0.135s} for BM25 and \textbf{2.84s} for SBERT -- further reduced to \textbf{0.065s} and \textbf{1.37s}, respectively, under parallel retrieval.
% Overall, generation dominates the runtime, yet ReDI remains practically efficient for reasoning-enhanced query understanding without task-specific optimization.
\begin{figure}[h]
  \centering
  \begin{tikzpicture}
    \begin{axis}[
      width=0.9\columnwidth,
      height=0.6\columnwidth,
      xlabel={Retrieval Unit Len},
      xlabel style={font=\fontsize{7pt}{6pt}\selectfont,yshift=2pt},
      ylabel={Average NDCG@10},
      ylabel style={font=\fontsize{7pt}{6pt}\selectfont,yshift=-4pt},
      ymin=18, ymax=32,
      % ymajorgrids,
      % grid style={white!70!black},
      xtick={3,5,7,9,11,13,15},
      xticklabels={3,5,7,9,11,13,15},
      yticklabel style={font=\fontsize{7pt}{6pt}\selectfont},
      xticklabel style={font=\fontsize{7pt}{6pt}\selectfont},
      clip=false,                             % 允许超出轴框的内容
      legend style={
        at={(0.5,1)},
        anchor=south,
        legend columns=4,
        font=\scriptsize,
        draw=none
      }
    ]
      % ——— Sparse 主折线 ———
      \addplot+[
        MidnightBlue, thick, mark=*, dotted,
        mark options={scale=1, fill=MidnightBlue, draw=MidnightBlue!40},
        every mark/.append style={solid},
        nodes near coords,
        nodes near coords style={anchor=south,yshift=2pt,font=\scriptsize}
      ] coordinates {
        (3,26.67) (5,27.14) (7,27.53) (9,27.68)
        (11,27.53) (13,27.46) (15,26.99)
      };
      \addlegendentry{Fixed-Sparse}

      % ——— Dense 主折线 ———
      \addplot+[
        Orange, thick, mark=triangle*, dotted,
        mark options={scale=1.2, fill=Orange, draw=Orange!40},
        every mark/.append style={solid},
        nodes near coords,
        nodes near coords style={anchor=south,yshift=2pt,font=\scriptsize}
      ] coordinates {
        (3,18.98) (5,19.95) (7,19.99) (9,20.61)
        (11,20.54) (13,20.18) (15,19.36)
      };
      \addlegendentry{Fixed-Dense}

      % ——— Flexible-Sparse 蓝色水平虚线 ———
      \addplot[
        MidnightBlue, solid, thick,
        domain=3:15, samples=2
      ] {30.8};
      \addlegendentry{Flexible-Sparse}
      % 在虚线右侧显示“30.8”
      \node[MidnightBlue,anchor=west,font=\scriptsize]
        at (axis cs:15,30.8) {30.8};
        
      % ——— Flexible-Dense 橙色水平虚线 ———
      \addplot[
        Orange, solid, thick,
        domain=3:15, samples=2
      ] {22.8};
      \addlegendentry{Flexible-Dense}
      % 在虚线右侧显示“22.8”
      \node[Orange,anchor=west,font=\scriptsize]
        at (axis cs:15,22.8) {22.8};

    \end{axis}
  \end{tikzpicture}
  \caption{nDCG@10 on BRIGHT with different numbers of sub‐queries and interpretation units.}
  \label{fig:unit-len}
\end{figure}

\subsection{Impact of strategy optimization}
\label{sec:ablation_finetune}
% Beyond module design, we explore how different training and retrieval strategies influence ReDI's effectiveness.

% Ablation Part 2.1 k3/desc weight
\heading{Hyperparameter sensitivity}
We analyze how retrieval performance responds to key hyperparameters in both sparse and dense settings. For \textbf{sparse retrieval} (Figure~\ref{fig:k3-singleplot}), we vary the $k_3$ parameter, which controls query-side term frequency scaling. On shorter documents (the blue curve), smaller $k_3$ values (0.2–0.8) yield better results, peaking at $k_3=0.4$ with an nDCG@10 of 38.25. In contrast, for longer documents (the orange curve), Recall@1 improves with larger $k_3$, reaching its maximum (25.98) at $k_3=5$ and plateauing thereafter. This suggests that shorter documents benefit from lower $k_3$, which emphasizes matching a wider variety of query terms, while longer documents require higher $k_3$ to strengthen core term signals within more expansive content. Beyond $k_3 = 5$, further increases yield diminishing returns. For \textbf{dense retrieval} (Figure~\ref{fig:desc-weight-single}), we vary the interpolation weight between the sub-query and its interpretation. nDCG@10 peaks at $\lambda=0.5$ (23.67), while Recall@1 reaches its maximum at $\lambda=0.4$ (23.12). Performance consistently drops as the interpolation shifts toward either extreme, highlighting the importance of balancing intent (sub-query) and contextual cues (interpretation). Further study of the complementarity between their embedding spaces is an interesting future direction.

%It would be interesting to further study the the embedding space
%Overweighting one component undermines the complementary strengths of the other.
\begin{figure}[h]
  \centering
  % ---- 统一 legend 部分 ----
  \begin{tikzpicture}
    \begin{axis}[
      hide axis,
      xmin=0, xmax=1, ymin=0, ymax=1,
      legend style={
        at={(0.5,1.1)},  % 控制图例整体位置
        anchor=south,
        legend columns=3,
        font=\scriptsize,
        draw=none
      }
    ]
      \addlegendimage{MidnightBlue, thick, mark=*, mark options={scale=1.1, fill=MidnightBlue, draw=MidnightBlue!40}}
      \addlegendentry{StackExchange NDCG@10}

      \addlegendimage{orange, thick, mark=triangle*, mark options={scale=1.5, fill=orange, draw=orange!40}}
      \addlegendentry{Long-doc Recall@1}
      \addlegendimage{red, thick, only marks, mark=x, mark options={scale=1.5, line width=1pt, fill=red, draw=red},}
      \addlegendentry{Max Point}
    \end{axis}
  \end{tikzpicture}
  \makebox[1\columnwidth][c]{
  % ---- 子图 1 ----
  \begin{subfigure}[t]{0.7\columnwidth}
    \captionsetup{skip=-2pt}
    \centering
    \begin{tikzpicture}
      \begin{axis}[
        width=0.9\linewidth,
        height=0.7\linewidth,
        xlabel={$k_3$},
        ymin=24, ymax=41,
        clip=false,
        symbolic x coords={0.2,0.4,0.8,2,5,10,20,50},
        xtick=data,
        xlabel style={font=\fontsize{7pt}{6pt}\selectfont,yshift=2pt},
        yticklabel style={font=\fontsize{6pt}{6pt}\selectfont},
        xticklabel style={font=\fontsize{7pt}{6pt}\selectfont},
      ]
        \addplot [MidnightBlue, thick] coordinates {
          (0.2,38.20) (0.4,38.25) (0.8,38.06) (2,36.80)
          (5,36.01)  (10,35.30) (20,35.08) (50,34.82)
        };
        \addplot+[
          only marks,
          mark=*, 
          mark options={scale=1.1, fill=MidnightBlue, draw=MidnightBlue!40},
        ] coordinates {
          (0.2,38.20) (0.4,38.25) (0.8,38.06) (2,36.80)
          (5,36.01)  (10,35.30) (20,35.08) (50,34.82)
        };
        \addplot+[
          only marks,
          mark=x, 
          mark options={scale=1.5, line width=1pt, fill=red, draw=red},
          nodes near coords,
          nodes near coords style={anchor=north,yshift=-2pt,font=\scriptsize},
          text=MidnightBlue
        ] coordinates {
          % 这里手动列出要标注的点（对应索引 0,2,4,6）
          (0.4,38.25)
        };

        \addplot [orange, thick] coordinates {
          (0.2,24.75) (0.4,25.03) (0.8,25.50) (2,25.96)
          (5,25.98)  (10,25.32) (20,25.78) (50,25.92)
        };
        \addplot+[
          only marks,
          mark=triangle*, 
          mark options={scale=1.5, fill=orange, draw=orange!40},
        ] coordinates {
          (0.2,24.75) (0.4,25.03) (0.8,25.50) (2,25.96)
          (5,25.98)  (10,25.32) (20,25.78) (50,25.92)
        };
        \addplot+[
          only marks,
          mark=x, 
          mark options={scale=1.5, line width=1pt, fill=red, draw=red},
          nodes near coords,
          nodes near coords style={anchor=south,yshift=2pt,font=\scriptsize},
          text=orange
        ] coordinates {
          % 这里手动列出要标注的点（对应索引 0,2,4,6）
          (5,25.98)
        };
      \end{axis}
    \end{tikzpicture}
    \caption{Sparse}
    \label{fig:k3-singleplot}
  \end{subfigure}
  \hspace{-5.5em}
  % ---- 子图 2 ----
  \begin{subfigure}[t]{0.7\columnwidth}
    \captionsetup{skip=-2pt}
    \centering
    \begin{tikzpicture}
      \begin{axis}[
        width=0.9\linewidth,
        height=0.7\linewidth,
        xlabel={Interpretation Weight},
        xlabel style={font=\fontsize{7pt}{6pt}\selectfont,yshift=2pt},
        ylabel style={font=\fontsize{7pt}{6pt}\selectfont,yshift=4pt},
        yticklabel style={font=\fontsize{6pt}{6pt}\selectfont},
        xticklabel style={font=\fontsize{7pt}{6pt}\selectfont},
        ymin=20.5, ymax=24.5,
        % ymajorgrids,
        % grid style={white!70!black},
        symbolic x coords={0.1,0.2,0.3,0.4,0.5,0.6,0.7,0.8,0.9},
        xtick={0.1,0.3,0.5,0.7,0.9},
        cycle list={
          {MidnightBlue, thick, mark=*, mark options={scale=1.1, fill=MidnightBlue, draw=MidnightBlue!40}},
          {orange, thick, mark=*, mark options={scale=1.2, fill=orange, draw=orange!40}},
        }
      ]
        \addplot [MidnightBlue, thick] coordinates {
          (0.1,21.31) (0.2,21.93) (0.3,22.76) (0.4,23.19)
        (0.5,23.67) (0.6,23.40) (0.7,22.73) (0.8,22.36)
        (0.9,21.42)
        };
        \addplot+[
          only marks,
          mark=*, 
          mark options={scale=1.1, fill=MidnightBlue, draw=MidnightBlue!40},
        ] coordinates {
          (0.1,21.31) (0.2,21.93) (0.3,22.76) (0.4,23.19)
        (0.5,23.67) (0.6,23.40) (0.7,22.73) (0.8,22.36)
        (0.9,21.42)
        };
        \addplot+[
          only marks,
          mark=x, 
          mark options={scale=1.5, line width=1pt, fill=red, draw=red},
          nodes near coords,
          nodes near coords style={anchor=south,yshift=2pt,font=\scriptsize},
          text=MidnightBlue
        ] coordinates {
          % 这里手动列出要标注的点（对应索引 0,2,4,6）
          (0.5,23.67)
        };

        \addplot [orange, thick] coordinates {
          (0.1,22.48) (0.2,22.81) (0.3,22.97) (0.4,23.12)
        (0.5,22.98) (0.6,22.36) (0.7,22.05) (0.8,21.60)
        (0.9,21.39)
        };
        \addplot+[
          only marks,
          mark=triangle*, 
          mark options={scale=1.5, fill=orange, draw=orange!40},
        ] coordinates {
          (0.1,22.48) (0.2,22.81) (0.3,22.97) (0.4,23.12)
        (0.5,22.98) (0.6,22.36) (0.7,22.05) (0.8,21.60)
        (0.9,21.39)
        };
        \addplot+[
          only marks,
          mark=x, 
          mark options={scale=1.5, line width=1pt, fill=red, draw=red},
          nodes near coords,
          nodes near coords style={anchor=north,yshift=-2pt,font=\scriptsize},
          text=orange
        ] coordinates {
          % 这里手动列出要标注的点（对应索引 0,2,4,6）
          (0.4,23.12)
        };
      \end{axis}
    \end{tikzpicture}
    \caption{Dense}
    \label{fig:desc-weight-single}
  \end{subfigure}
  }
  \caption{Performance of the sparse retriever under different $k_3$ and dense retriever at varying interpretation weights.}
  \label{fig:k3&weight}
\end{figure}

% Ablation Part 2.2 Joint vs Two
\heading{Fine-tuning method}
We compare joint fine-tuning with two-stage fine-tuning (as detailed in Section~\ref{sec:model-fine-tuning}) to study the impact of the learning strategy. As shown in Table \ref{tab:joint_vs_twostage}, the two-stage training consistently outperforms joint training across both retrieval settings. Specifically, for sparse retrieval, it yields improvements of $8.2\%$ in nDCG@10 on StackExchange tasks and $8.8\%$ on overall tasks, while for dense retrieval, the gains reach $8.7\%$ and $9.6\%$, respectively. These results demonstrate the advantage of decoupling learning objectives, which enable each stage to specialize without gradient interference, resulting in improved stability and overall retrieval effectiveness.

\begin{table}[H]
  \centering
  \caption{nDCG@10 on BRIGHT: Joint vs.\ two-stage training}
  \label{tab:joint_vs_twostage}
  \small
  \setlength{\tabcolsep}{5pt}
  \begin{tabular}{
                  l   % Retriever
                  l   % Query Rewrite Model
                  cccc}
    \toprule
    Retriever & Model & StackExchange & Coding & Theo. & \textbf{Avg.All}\\
    \midrule
    \multirow{2}{*}{BM25}   
    & Joint                        & 35.4 & 15.3 & 20.6 & 28.3 \\
    & Two-Stage                    & \textbf{38.3} & \textbf{17.3} & \textbf{22.5} & \textbf{30.8} \\

    \midrule
    \multirow{2}{*}{SBERT}
    & Joint                        & 21.8 & 19.4 & 19.1 & 20.8 \\
    & Two-Stage                    & \textbf{23.7} & \textbf{21.2} & \textbf{21.7} & \textbf{22.8} \\ 
    \bottomrule
  \end{tabular}
\end{table}

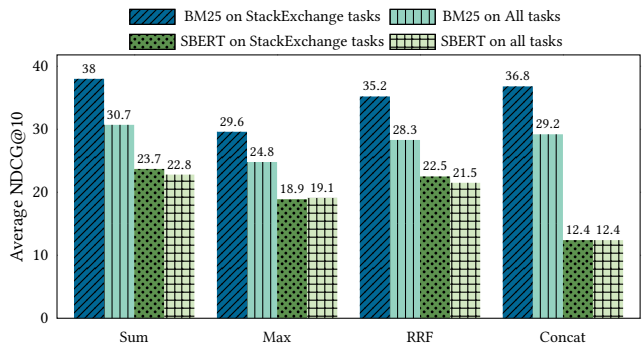
\begin{figure}[!t]
  \centering
  \makebox[1\columnwidth][c]{
  \begin{tikzpicture}
    % -------- Legend Axis --------
    \begin{axis}[
      hide axis,
      xmin=0, xmax=1, ymin=0, ymax=1,
      scale only axis,
      height=0pt, width=0pt,
      legend style={
        at={(10,-10)},
        anchor=north,
        legend columns=2,
        font=\scriptsize,
        draw=none,
      }
    ]
      \addlegendimage{area legend, fill=MidnightBlue, postaction={pattern=north east lines}}
      \addlegendentry{BM25 on StackExchange tasks}
      \addlegendimage{area legend, fill=SeaGreen!60, postaction={pattern=vertical lines}}
      \addlegendentry{BM25 on All tasks}
      \addlegendimage{area legend, fill=OliveGreen!80, postaction={pattern=crosshatch dots}}
      \addlegendentry{SBERT on StackExchange tasks}
      \addlegendimage{area legend, fill=YellowGreen!40, postaction={pattern=grid}}
      \addlegendentry{SBERT on all tasks}
    \end{axis}
    % -------- Bar Plot Axis --------
    \begin{axis}[
      at={(0,-0.75cm)},  % 控制主图与 legend 的垂直距离
      anchor=north,
      ybar,
      axis on top,
      height=0.6\linewidth,
      width=0.95\linewidth,
      x=1.9cm,
      bar width=0.4cm,
      enlarge x limits=0.18,
      tickwidth=0pt,
      symbolic x coords={Sum,Max,RRF,Concat},
      xtick=data,
      ymin=0,
      ylabel={Average NDCG@10},
      ylabel style={font=\fontsize{7pt}{6pt}\selectfont,yshift=-4pt},
      xticklabel style={
        anchor=north,
        yshift=-2pt,
        font=\fontsize{6pt}{6pt}\selectfont
      },
      yticklabel style={
        font=\fontsize{6pt}{6pt}\selectfont
      },
    ]
      \addplot[
        draw=none,
        fill=MidnightBlue,
        postaction={pattern=north east lines},
        bar shift=-0.6cm,
        nodes near coords,
        nodes near coords style={font=\fontsize{5.5pt}{4pt}\selectfont, yshift=-1pt},
      ] coordinates {
        (Sum,38.0) (Max,29.6) (RRF,35.2) (Concat,36.8)
      };

      \addplot[
        draw=none,
        fill=SeaGreen!60,
        postaction={pattern=vertical lines},
        bar shift=-0.2cm,
        nodes near coords,
        nodes near coords style={font=\fontsize{5.5pt}{4pt}\selectfont, yshift=-1pt},
      ] coordinates {
        (Sum,30.7) (Max,24.8) (RRF,28.3) (Concat,29.2)
      };

      \addplot[
        draw=none,
        fill=OliveGreen!80,
        postaction={pattern=crosshatch dots},
        bar shift=+0.2cm,
        nodes near coords,
        nodes near coords style={font=\fontsize{5.5pt}{4pt}\selectfont, yshift=-1pt},
      ] coordinates {
        (Sum,23.7) (Max,18.9) (RRF,22.5) (Concat,12.4)
      };

      \addplot[
        draw=none,
        fill=YellowGreen!40,
        postaction={pattern=grid},
        bar shift=+0.6cm,
        nodes near coords,
        nodes near coords style={font=\fontsize{5.5pt}{4pt}\selectfont, yshift=-1pt},
      ] coordinates {
        (Sum,22.8) (Max,19.1) (RRF,21.5) (Concat,12.4)
      };
    \end{axis}
  \end{tikzpicture}
  }
  \caption{nDCG@10 on BRIGHT with different retrieval fusion methods.}
  \label{fig:fusion_method}
\end{figure}

\heading{Fusion methods}
We compare four strategies for aggregating retrieval results across units: score summation (sum), highest score (max), reciprocal rank fusion (RRF), and single merged query (concat). As shown in Figure~\ref{fig:fusion_method}, sum fusion consistently delivers the best performance. While concat performs comparably to sum in sparse retrieval, its performance drops sharply in dense retrieval, indicating that long merged queries dilute semantic focus and confuse dense encoders. RRF and max yield moderate or lower results across all settings. These results highlight the robustness of score-based aggregation, particularly in dense retrieval, where preserving unit granularity is crucial for maintaining semantic precision.

% Ablation Part 3 Transferability evaluation
\subsection{Transferability and generalization}
\label{sec:ablation_trans}
\heading{Long document retrieval}
Table~\ref{tab:main_result_long} reports the performance on the BRIGHT StackExchange long-document subset. 
In general, ReDI surpasses all reasoning-expanded and feedback-based baselines over the average Recall@1 for both sparse and dense retrieval. It achieves 26.0\% in the sparse setting, leading 6 of 7 tasks, and 23.1\% in the dense setting, ranking first on 3 of 7 tasks.
These results highlight ReDI's strong generalization to long documents and validate the effectiveness of our reasoning decomposition with interpretation strategy.

% already defined in definitions.tex
\if0
\newcommand{\sig}{\rlap{$^{*}$}}
\newcommand{\num}[1]{$#1$}
\newcommand{\nums}[1]{$#1$\sig}
\newcommand{\numb}[1]{$\mathbf{#1}$}
\newcommand{\numbs}[1]{$\mathbf{#1}$\sig}
\newcommand{\numu}[1]{$\underline{#1}$}
\newcommand{\numus}[1]{$\underline{#1}$\sig}
\fi

\begin{table}[h]
  \centering
  \caption{Recall@1 on the BRIGHT StackExchange long‐document subset.}
  \label{tab:main_result_long}
  %\small
    \renewcommand{\arraystretch}{0.9}

  \setlength{\tabcolsep}{2pt}
  \begin{adjustbox}{max width=\columnwidth}
  \begin{tabular}{
                  % l   % Retriever
                  l   % Query Rewrite Model
                  rrrrrrrr
                  % cccccccc
                  }
    \toprule
    % \multirow{2}{*}{\numb{Retriever}}
      \multirow{2}{*}{\numb{Model}}
      & \multicolumn{8}{c}{\textit{StackExchange}}\\
    \cmidrule(lr){2-9}
      & Bio. & Earth. & Econ. & Psy. & Rob. & Stack. & Sus. & \numb{Avg.}\\
    \midrule
    %\rowcolor{gray!20} 
    \multicolumn{9}{c}{\emph{Using BM25 Retriever}}\\
    \midrule
     -             & \num{10.7} & \num{15.4} & \num{10.7} & \num{8.4} & \num{7.4} & \num{22.2} & \num{10.7} & \num{12.2} \\
     \midrule
     Claude-3-opus & 26.8 & \num{13.5} & 13.4 & 28.2 & \num{7.9} & 28.2 & \num{11.8} & \num{18.5} \\
     GPT4          & 26.8 & \num{15.8} & \num{10.2} & 30.7 & \num{5.9} & \num{26.5} & \num{9.7} & \num{17.9} \\
     DeepSeek-R1   & 26.8 & 20.0 & 14.4 & 30.2 & 14.9 & 33.3 & \num{10.6} & \num{21.5} \\
     \midrule
     TongSearchQR    & 29.4 & 17.8 & 15.4 & 27.1 & 10.4 & 26.5 & 12.0 & 19.8 \\
     DIVER-QExpand         & \numb{34.4} & 21.7 & 16.0 & 27.2 & \num{16.3} & \num{29.5} & \num{21.1} & \num{23.7} \\
     \textbf{ReDI} & 28.4 & \numb{22.4} & \numb{21.2} & \numb{32.0} & \numb{19.8} & \numb{36.3} & \numb{21.7} & \numb{26.0} \\

     \midrule
     %\rowcolor{gray!20} 
     \multicolumn{9}{c}{\emph{Using SBERT Retriever}}\\
     \midrule
     -             & \num{25.6} & 34.1 & 18.9 & 15.8 & 10.9 & 15.0 & 18.0 & \num{19.7} \\
     \midrule
     Claude-3-opus & 34.8 & 31.6 & 21.8 & 15.8 & 8.9 & 15.8 & 16.6 & \num{20.8} \\
     GPT4          & 37.7 & \numb{35.3} & 19.9 & 18.3 & 12.4 & 11.5 & \numb{22.6} & 22.5 \\
     DeepSeek-R1   & 35.6 & 34.8 & 16.0 & 15.3 & 8.9 & 15.0 & 19.9 & \num{20.8} \\
     \midrule
     TongSearchQR    & 31.4 & 33.8 & \numb{22.8} & 14.9 & \num{13.9} & 13.2 & 18.5 & 21.2 \\
     DIVER-QExpand         & \numb{40.1} & 35.0 & 18.0 & 16.8 & 12.9 & 15.0 & 20.3 & 22.6 \\
     \textbf{ReDI} & 36.2 & 32.8 & \numb{22.8} & \numb{20.8} & 10.9 & \numb{16.2}  & 22.2 & \numb{23.1} \\
    \bottomrule
  \end{tabular}
  \end{adjustbox}
\end{table}

\heading{Out-of-domain retrieval}
Finally, we examine the generalization ability of ReDI by evaluating it on shorter queries from the BEIR benchmark. As is shown in Table~\ref {tab:beir_results}, ReDI with BM25 achieves an average nDCG@10 of 44.9 across nine tasks, surpassing Rank1-7B (40.9), MonoT5-3B (44.7), and RankLLaMA-7B (44.4). ReDI ranks among the top systems on multiple tasks, demonstrating strong out-of-domain generalization and confirming the effectiveness of our structured decomposition and interpretation model for real-world retrieval beyond BRIGHT.
\begin{table}[!t]
  \centering
  \caption{nDCG@10 on BEIR. $^\dagger$Results from~\citet{weller2025rank1}. Significant improvement or degradation with respect to ReDI is indicated (+/-) (p-value $\leq$ 0.05).}
  \label{tab:beir_results}
  \small
  \setlength{\tabcolsep}{0pt}      
  % \resizebox{\columnwidth}{!}{%
    \begin{tabular}{%
      l % Query Rewrite Model
      cccccccccc}
      \toprule
      Model & ArguA. & ClimF. & DBP. & FiQA. & NFC. & SciD. & SciF. & Touche. & TrecC. & \textbf{Avg.} \\
      \midrule
      BM25 Flat               & \numm{39.7} & \numm{16.5} & \numm{31.8} & \numm{23.6} & \numm{32.2}  & \numm{14.9} & \numm{67.9}  & \numum{44.2}  & \numm{59.5}  & \numm{36.7} \\
      
      BM25S                   & \numup{48.3} & \numm{18.1} & \numm{27.0} & \numm{23.3} & \numm{30.6} & \numm{15.4} & \numm{66.2} & \numm{35.8} & \numm{63.5} & \numm{36.5} \\
      
      $\text{+\textbf{ReDI}}$ & 44.7 & \textbf{29.5} & 42.0 & 26.3 & \textbf{39.4} & \numu{18.0} & 74.5 & \textbf{49.3} & \numu{80.7} & \textbf{44.9} \\
      
      \midrule
      
      $\text{MonoT5-3B}^\dagger$    & \numm{42.5} & \numum{25.4} & \numbp{44.5} & \numbp{46.5} & \numu{37.8}  & \textbf{19.3} & \numu{76.1}  & \numm{30.7}  & 79.6  & \numum{44.7} \\
      $\text{RankLLaMA-7B}^\dagger$ & \numbp{54.4} & \numm{23.2} & \numu{43.7} & \numup{42.1} & \numm{27.0} & 16.6 & \numm{71.1} & \numm{41.4}  & 80.2  & \numm{44.4} \\
      $\text{Rank1-7B}^\dagger$     & \numm{42.8} & \numm{15.0} & \numm{38.9} & \nump{39.5} & \numm{36.2}  & 17.2 & \numbp{77.2}  & \numm{22.8}  & \textbf{81.9}  & \numm{40.9} \\
      \bottomrule
    \end{tabular}%
  % } % end \resizebox
\end{table}

\subsection{Interaction with reasoning-oriented retriever}
\label{sec:reasonir}
Table~\ref{tab:reasonir_results} shows that ReDI, when paired with the reasoning-native retriever ReasonIR-8B,\footnote{\url{https://huggingface.co/reasonir/ReasonIR-8B}} yields an average score of $30.7$ across all tasks and $32.7$ on StackExchange, underperforming single long-form expansion methods. While initially counter-intuitive, we attribute this to a conflict between reasoning modalities. ReasonIR, fine-tuned on long synthetic chain-of-thought data, performs implicit decomposition within its high-dimensional latent space (4096 dimensions). ReDI's explicit fragmentation inadvertently disrupts these holistic semantic signals, leading to suboptimal alignment.

\noindent%
This suggests a dichotomy in retrieval paradigms: lightweight retrievers, such as BM25 and SBERT, lack intrinsic reasoning capacity and therefore benefit substantially from an explicit \textbf{reasoning-then-retrieving} strategy. In contrast, heavy reasoning-retrievers like ReasonIR are suited to a \textbf{retrieving-by-reasoning} paradigm, where complex intent is preserved in a single representation.

\begin{table}[H]
  \centering
    %\small
  \caption{nDCG@10 on BRIGHT with ReasonIR-8B.}
  \label{tab:reasonir_results}
  \renewcommand{\arraystretch}{0.9}
  \setlength{\tabcolsep}{3pt}
  \begin{tabular}{l c c c c}
    \toprule
    Model & StackExchange & Coding & Theo. & \textbf{Avg. all} \\
    \midrule
    % %\rowcolor{gray!20} 
    % \multicolumn{5}{c}{\emph{Using ReasonIR Retriever}}\\
    % \midrule
    GPT-4 Reason-query & 31.2 & \num{24.9} & 26.5 & \num{28.8} \\
    TongSearch-QR      & \num{34.2} & \num{24.2} & \numb{30.1} & 31.8 \\
    DIVER-QExpand      & \numb{35.9} & \num{21.3} & 29.1 & \numb{32.2} \\
    \midrule    
    \numb{ReDI}        & 32.7 & \numb{28.5} & 27.4 & 30.7 \\
    % \midrule
    % %\rowcolor{gray!20} 
    % \multicolumn{5}{c}{\emph{Using DIVER Retriever}}\\
    % \midrule
    % GPT-4 Reason-query & \num{36.2} & 20.2 & 30.0 & \num{32.0} \\
    % TongSearch-QR      & \num{32.3} & 16.8 & 31.5 & \num{29.5} \\
    % DIVER-QExpand      & \num{36.0} & 21.1 & \numbm{33.5} & \num{32.9} \\
    % \midrule    
    % \numb{ReDI}        & \numb{39.0} & \numb{21.2} & 31.4 & \numb{34.2} \\
    \bottomrule
  \end{tabular}
  \medskip
\end{table}

This distinction defines ReDI's practical value. 
As shown in Table~\ref{tab:main_result_1}, \textit{BM25+ReDI} achieves an average score of $38.3$ on StackExchange tasks, surpassing the strongest baseline \textit{ReasonIR+DIVER-QExpand} ($36.4$), while relying on a substantially lighter retriever. 
Furthermore, ReDI demonstrates significant efficiency advantages: as reported in Table~\ref{tab:efficiency}, \textit{BM25+ReDI} and \textit{SBERT+ReDI} are $58\times$ and $4\times$ faster than \textit{ReasonIR+Qwen3-8B} during retrieval, resulting in a $1.6\times$ improvement in overall efficiency.

These results indicate that ReDI is best positioned as a ``plug-and-play'' reasoning module that elevates standard retrieval systems to reasoning-capable performance, without incurring the computational cost of deploying heavy reasoning-native retrievers. This makes ReDI particularly attractive for scalable, real-world retrieval deployments.

\begin{table}[H]
\centering
\caption{Runtime of different configurations.}
\label{tab:efficiency}
% \small
\setlength{\tabcolsep}{3pt}
\begin{tabular}{lllcc}
\toprule
QU Model & Retriever & Setting & Avg./Q (s) & Total(s) \\
\midrule
\multirow{2}{*}{ReDI} 
 & \multirow{2}{*}{BM25} 
 & Decomp. + Interp. & 8.50 & \multirow{2}{*}{8.64} \\
 &  & Retrieve & 0.14 &  \\[1pt]
\midrule
\multirow{2}{*}{ReDI} 
 & \multirow{2}{*}{SBERT}
 & Decomp. + Interp. & 7.74 & \multirow{2}{*}{9.89} \\
 &  & Retrieve & 2.15 &  \\[1pt]
\midrule
\multirow{2}{*}{Qwen3-8B} 
 & \multirow{2}{*}{ReasonIR}
 & Single long-form & 5.34 & \multirow{2}{*}{13.50} \\
 &  & Retrieve & 8.16 &  \\[1pt]
\midrule
\multirow{2}{*}{DeepSeek-R1} 
 & \multirow{2}{*}{ReasonIR}
 & Single long-form & 13.50 & \multirow{2}{*}{30.27} \\
 &  & Retrieve & 16.77 &  \\
\bottomrule
\end{tabular}
\end{table}

\section{Conclusion}
We have proposed \textbf{ReDI}, a reasoning-enhanced model for complex query understanding (QU) that addresses the core challenge of faithfully aligning a user’s multi-faceted information need with retrievable evidence. By explicitly decomposing each complex query into targeted sub-queries and augmenting them with concise, intent-preserving interpretations, our modular pipeline enables unit-level retrieval followed by principled score fusion. Experiments on the BRIGHT and BEIR benchmarks confirm that this design substantially improves retrieval effectiveness across both sparse and dense paradigms.

While ReDI is effective, several limitations highlight directions for future work. First, the improvements under dense retrieval are less pronounced than those under sparse retrieval, suggesting a potential mismatch between dense representations and fine-grained query semantics. Second, the current decomposition heavily relies on the \ac{LLM}’s internal knowledge, developing more controlled methods to decide \textit{what} and \textit{when} to decompose remains an intriguing avenue for exploration.
% leverage external resources to control 
% incorporating external signals -- such as graph structures, user click trails, or shallow Web snippets -- could guide more robust sub-query generation, especially in knowledge-sparse domains. 
Third, free-form interpretations may introduce spurious semantics that degrade retrieval accuracy, motivating future work on controllable generation, factuality constraints, and retrieval-grounded verification. Finally, while ReDI is competitive, ReasonIR’s results suggest that its gains are less stable across domains, motivating future work on retriever-adaptive interpretations to further enhance robustness.
Addressing these limitations would enhance both the generality and robustness of reasoning-based \acl{QU}, paving the way for broader adoption in real-world tasks such as complex open-domain QA, conversational agents, and personalized search.

%\input{figures/equations}

% \begin{acks}

% \end{acks}


% \clearpage
\bibliographystyle{ACM-Reference-Format}
\balance
\bibliography{references}

\clearpage
\appendix
\section{Appendix}
\label{sec:appendix}
% \subsection{Complete Experimental Results}

\heading{Role of Model Reasoning}
Table~\ref{tab:model size full} presents the complete experimental results in Figure~\ref{fig:model_size}. Across BM25, scaling the rewrite model and enabling “think” generally increase nDCG@10; Qwen3-8B(think) gives the largest gains among Qwen variants. Under SBERT, gains from “think” are smaller and sometimes negative for small models, suggesting dense encoders are less tolerant of verbose/latent-logic rewrites. In both retrievers, \textbf{ReDI} surpasses all size/think settings, indicating that structured decomposition + interpretation contributes more than raw model size or implicit chain-of-thought.

\begin{table*}[!t]
 \centering
 \caption{nDCG@10 on BRIGHT with Qwen3 across different model sizes and reasoning models.}
 \label{tab:model size full}
 %\small
 \setlength{\tabcolsep}{7pt}
 \begin{tabular}{%
  l % Query Rewrite Model Params
  l % Think Mode
  c % Avg.All (formerly Avg.)
  c % Avg.ES (formerly StackExchange/Avg.)
  r@{\;\,}r@{\;\,}r@{\;\,}r@{\;\,}r@{\;\,}r@{\;\,}r
  r@{\;\,}r
  r@{\;\,}r@{\;\,}r}
  \toprule    
   \multirow{2}{*}{\textbf{Param}}
   & \multirow{2}{*}{\textbf{Think Mode}}
   & \multirow{2}{*}{\textbf{Avg.All}}
   & \multirow{2}{*}{\textbf{Avg.SE}}
   & \multicolumn{7}{c}{\textit{StackExchange}}
   & \multicolumn{2}{c}{\textit{Coding}}
   & \multicolumn{3}{c}{\textit{Theorem‐based}} \\
  \cmidrule(lr){5-11} \cmidrule(lr){12-13} \cmidrule(lr){14-16}
   &
   &
   &
   & Bio. & Earth. & Econ. & Psy. & Rob. & Stack. & Sus.
   & Leet. & Pony
   & AoPS & TheoQ. & TheoT. \\
  \midrule
   %\rowcolor{gray!20} 
   \multicolumn{16}{c}{\emph{Using BM25 Retriever}}\\
   \midrule
   - & -                & 14.5 & 17.2 & 18.9 & 27.2 & 14.9 & 12.5 & 13.6 & 18.4 & 15.0 & \textbf{24.4} & \textbf{7.9} & \textbf{6.2} & 10.4 & 4.9 \\
   \midrule
   0.6B & Disabled      & 15.3 & 19.7 & 22.0 & 31.9 & 17.0 & 23.9 & 13.9 & 16.6 & 12.7 & 14.4 & 3.4 & 1.1 & 15.5 & 11.3 \\
   0.6B & Enabled       & 15.7 & 19.3 & 26.7 & 26.2 & 15.3 & 21.6 & 13.5 & 14.6 & 17.0 & 20.6 & 6.3 & 2.3 & 15.8 & 8.0  \\
   4B & Disabled        & 19.0 & 24.9 & 28.8 & 39.3 & 10.8 & 27.3 & 19.6 & 21.0 & 16.7 & 19.4 & 3.7 & 2.4 & 19.3 & 19.5 \\
   4B & Enabled         & 20.1 & 25.7 & 32.6 & 40.5 & 18.4 & 28.3 & 17.3 & 22.5 & \textbf{20.3} & 18.8 & 3.8 & 3.0 & 17.6 & 18.0 \\
   8B & Disabled        & 20.7 & 26.3 & 32.8 & 43.3 & 18.2 & \textbf{30.3} & 19.6 & 22.4 & 17.3 & 19.4 & 3.7 & 2.4 & 19.3 & 19.5 \\
   8B & Enabled         & \textbf{22.8} & \textbf{29.0} & \textbf{40.1} & \textbf{46.0} & \textbf{20.9} & 29.6 & \textbf{21.5} & \textbf{25.0} & 20.2 & 16.9 & 6.4 & 3.1 & \textbf{22.5} & \textbf{21.7} \\
   \midrule
   %\rowcolor{gray!20} 
   \multicolumn{16}{c}{\emph{Using SBERT Retriever}}\\
   \midrule
   - & -                & 14.9 & 15.6 & 15.1 & 20.4 & \textbf{16.6} & \textbf{22.7} & 8.2  & 11.0 & 15.3 & \textbf{26.4} & 7.0 & \textbf{5.3} & 20.0 & 10.8 \\
   \midrule
   0.6B & Disabled      & 10.7 & 12.6 & 12.4 & 19.0 & 11.8 & 19.1 & \textbf{10.0} & 6.8  & 9.1  & 12.3 & 0.7 & 3.5 & 14.3 & 9.1  \\
   0.6B & Enabled       & 10.3 & 12.2 & 13.5 & 18.2 & 13.9 & 3.3  & 9.1  & 9.1  & 18.3 & 9.5  & 4.0 & 3.6 & 15.7 & 5.1  \\
   4B & Disabled        & 13.7 & 14.4 & \textbf{17.9} & 18.3 & 13.5 & 20.4 & 7.6  & 10.9 & 12.4 & 15.8 & 3.5 & 5.4 & 19.1 & 19.7 \\
   4B & Enabled         & 14.3 & 15.0 & 15.1 & 20.8 & 12.0 & \textbf{22.7} & 5.4  & 12.3 & 16.5 & 18.0 & 6.1 & 1.8 & 23.0 & 17.6 \\
   8B & Disabled        & 15.3 & 16.6 & 17.5 & \textbf{24.4} & 14.4 & 21.9 & 6.5  & 10.8 & \textbf{20.7} & 18.9 & 10.8 & 1.3 & 22.7 & 13.9 \\
   8B & Enabled         & \textbf{16.8} & \textbf{17.2} & 15.7 & 23.0 & 15.7 & 22.1 & 9.3  & \textbf{14.9} & 19.8 & 18.2 & \textbf{11.2} & 3.4 & \textbf{26.9} & \textbf{21.2} \\
   \bottomrule
  \end{tabular}
\end{table*}

\heading{Flexible vs. Fixed Decomposition Granularity}
Table~\ref{tab:Unit compare full} presents the complete experimental results in Figure~\ref{fig:unit-len}. With BM25, performance improves from 3→9 units and plateaus after 9–11, while very large unit counts show mild regressions—evidence of coverage–noise trade-off. With SBERT, trends are flatter and peaks occur around 9–11, reflecting dense encoders’ preference for fewer, stronger facets. In both cases, \textbf{ReDI} (learned, intent-adaptive granularity) exceeds any fixed setting, supporting adaptive decomposition over manual heuristics.

\begin{table*}[!t]
  \centering
  \caption{nDCG@10 on BRIGHT with different nums of sub-query + interpretation unit.}
  \label{tab:Unit compare full}
  % \small
  \setlength{\tabcolsep}{11.5pt}
  \begin{tabular}{%
  c| % unit num
  c % Avg.All (formerly Avg.)
  c| % Avg.ES (formerly StackExchange/Avg.)
  r@{\;\,}r@{\;\,}r@{\;\,}r@{\;\,}r@{\;\,}r@{\;\,}r
  r@{\;\,}r
  r@{\;\,}r@{\;\,}r}
  \toprule    
   \multirow{2}{*}{\textbf{Num}}
   & \multirow{2}{*}{\textbf{Avg.All}}
   & \multirow{2}{*}{\textbf{Avg.SE}}
   & \multicolumn{7}{c}{\textit{StackExchange}}
   & \multicolumn{2}{c}{\textit{Coding}}
   & \multicolumn{3}{c}{\textit{Theorem‐based}} \\
  \cmidrule(lr){4-10} \cmidrule(lr){11-12} \cmidrule(lr){13-15}
   &
   &
   & Bio. & Earth. & Econ. & Psy. & Rob. & Stack. & Sus.
   & Leet. & Pony
   & AoPS & TheoQ. & TheoT. \\
    \midrule
    % \rowcolor{gray!20} 
    \multicolumn{15}{c}{\emph{Using BM25 Retriever}}\\
    \midrule  
    3     & 26.67 & 33.28 & 44.83 & 50.82 & 25.25 & 36.20 & 22.43 & 26.79 & 26.63 & 21.81 & 7.99 & 5.07 & 25.65 & 26.57 \\
    5     & 27.14 & 34.16 & 44.60 & 51.27 & 27.26 & 37.66 & 21.62 & 28.45 & 28.28 & 21.82 & 7.42 & 4.77 & 24.50 & 28.01 \\
    7     & 27.53 & 34.70 & 46.07 & 49.12 & 26.79 & 38.13 & 22.03 & 31.41 & 29.35 & 23.35 & 7.91 & 4.94 & 24.49 & 26.76 \\
    9     & 27.68 & 35.17 & 47.10 & 50.22 & 24.76 & 41.00 & 23.28 & 30.64 & 29.20 & 23.16 & 5.49 & 4.92 & 24.78 & 27.67 \\
    11    & 27.53 & 34.75 & 47.57 & 48.52 & 26.80 & 39.91 & 22.93 & 29.05 & 28.49 & 23.25 & 6.82 & 5.07 & 24.90 & 27.06 \\
    13    & 27.46 & 34.87 & 47.75 & 48.76 & 26.52 & 41.40 & 23.30 & 29.09 & 27.26 & 22.42 & 6.45 & 4.58 & 25.17 & 26.86 \\
    15    & 26.99 & 34.34 & 48.17 & 48.72 & 24.88 & 40.53 & 22.14 & 29.59 & 26.33 & 21.95 & 5.45 & 4.15 & 24.83 & 27.18 \\
    Flex. & 30.82 & 38.25 & 48.96 & 53.52 & 28.65 & 43.36 & 27.50 & 36.33 & 29.42 & 25.32 & 9.30 & 5.98 & 31.47 & 30.02 \\
    
    \midrule
    % \rowcolor{gray!20} 
    \multicolumn{15}{c}{\emph{Using SBERT Retriever}}\\
    \midrule 
    3     & 18.98 & 20.35 & 21.48 & 26.91 & 18.83 & 26.24 & 11.13 & 15.59 & 22.30 & 21.65 & 13.41 & 3.47 & 25.97 & 20.81 \\
    5     & 19.95 & 20.83 & 22.14 & 30.44 & 17.67 & 25.55 & 9.87  & 16.28 & 23.83 & 20.69 & 15.38 & 3.63 & 30.85 & 23.06 \\
    7     & 19.99 & 21.05 & 22.95 & 30.90 & 17.62 & 26.68 & 10.05 & 17.48 & 21.66 & 21.13 & 14.48 & 3.77 & 30.51 & 22.67 \\
    9     & 20.61 & 21.17 & 23.03 & 30.31 & 17.13 & 26.65 & 10.06 & 16.89 & 24.12 & 21.69 & 16.22 & 3.94 & 30.01 & 27.28 \\
    11    & 20.54 & 21.21 & 22.68 & 30.98 & 18.37 & 26.70 & 9.30  & 17.20 & 23.22 & 21.70 & 15.51 & 3.85 & 29.47 & 27.54 \\
    13    & 20.18 & 21.04 & 24.09 & 29.60 & 17.85 & 26.46 & 9.52  & 15.91 & 23.83 & 21.35 & 16.07 & 3.87 & 29.48 & 24.11 \\
    15    & 19.36 & 20.31 & 21.13 & 27.97 & 18.38 & 26.20 & 9.29  & 16.48 & 22.72 & 20.30 & 14.58 & 3.79 & 27.57 & 23.90 \\
    Flex. & 22.80 & 23.70 & 25.00 & 32.30 & 20.80 & 28.00 & 13.80 & 20.20 & 25.60 & 25.20 & 17.10 & 6.20 & 33.20 & 25.80 \\
    \bottomrule
  \end{tabular}
\end{table*}

\begin{table*}[!t]
  \centering
  \caption{Performance of sparse retriever (BM25) under different $k_3$.}
  \label{tab:bright-bm25tf-nDCG10}
  %\small
  \setlength{\tabcolsep}{12pt}
  \begin{tabular}{%
  c| % k_3
  c % Avg.All (formerly Avg.)
  c| % Avg.ES (formerly StackExchange/Avg.)
  r@{\;\,}r@{\;\,}r@{\;\,}r@{\;\,}r@{\;\,}r@{\;\,}r
  r@{\;\,}r
  r@{\;\,}r@{\;\,}r}
  \toprule    
   \multirow{2}{*}{$\mathbf{ k_3 }$}
   & \multirow{2}{*}{\textbf{Avg.All}}
   & \multirow{2}{*}{\textbf{Avg.SE}}
   & \multicolumn{7}{c}{\textit{StackExchange}}
   & \multicolumn{2}{c}{\textit{Coding}}
   & \multicolumn{3}{c}{\textit{Theorem‐based}} \\
  \cmidrule(lr){4-10} \cmidrule(lr){11-12} \cmidrule(lr){13-15}
   &
   &
   & Bio. & Earth. & Econ. & Psy. & Rob. & Stack. & Sus.
   & Leet. & Pony
   & AoPS & TheoQ. & TheoT. \\
   \midrule
    0    & 30.55 & 38.09 & 49.08 & 54.07 & 28.30 & 43.17 & 26.77 & 36.33 & 28.86 & 25.06 & 8.89 & 5.82 & 31.06 & 29.16 \\
    0.2  & 30.75 & 38.20 & 48.93 & 53.77 & 28.16 & 43.52 & 27.38 & 36.49 & 29.13 & 25.24 & 9.16 & 5.99 & 31.57 & 29.64 \\
    0.4  & 30.82 & 38.25 & 48.96 & 53.52 & 28.65 & 43.36 & 27.50 & 36.33 & 29.42 & 25.32 & 9.30 & 5.98 & 31.47 & 30.02 \\
    0.8  & 30.73 & 38.06 & 48.51 & 53.55 & 29.43 & 42.43 & 27.18 & 36.27 & 29.05 & 25.87 & 9.15 & 5.96 & 31.36 & 30.03 \\
    2    & 29.89 & 36.80 & 46.81 & 51.65 & 28.11 & 41.03 & 27.21 & 35.31 & 27.49 & 25.34 & 8.85 & 6.41 & 30.55 & 29.90 \\
    5    & 29.19 & 36.01 & 45.94 & 49.85 & 28.51 & 40.18 & 26.88 & 34.03 & 26.72 & 24.44 & 8.18 & 6.40 & 29.68 & 29.47 \\
    10   & 28.60 & 35.30 & 44.63 & 48.62 & 28.54 & 38.76 & 26.30 & 33.60 & 26.64 & 24.33 & 8.11 & 6.16 & 29.06 & 28.43 \\
    20   & 28.31 & 35.08 & 44.18 & 48.31 & 28.72 & 38.63 & 26.16 & 33.38 & 26.13 & 23.71 & 8.18 & 5.74 & 28.93 & 27.69 \\
    50   & 28.14 & 34.82 & 43.81 & 48.06 & 28.76 & 38.01 & 25.62 & 33.48 & 26.03 & 23.62 & 8.00 & 5.66 & 29.03 & 27.56 \\
    80   & 28.09 & 34.79 & 43.61 & 48.05 & 28.81 & 37.99 & 25.55 & 33.47 & 26.02 & 23.51 & 7.95 & 5.62 & 28.96 & 27.56 \\
    100  & 28.06 & 34.77 & 43.60 & 48.05 & 28.73 & 37.97 & 25.55 & 33.47 & 26.03 & 23.51 & 7.95 & 5.62 & 28.96 & 27.26 \\
    \bottomrule
  \end{tabular}
\end{table*}

\heading{Hyperparameter Sensitivity}
Table~\ref{tab:bright-bm25tf-nDCG10} presents the complete experimental results in Figure~\ref{fig:k3-singleplot}. BM25 attains its best average around $k_3 \approx 0.4$, with clear degradation as $k_3$ increases (i.e., approaching linear qtf). Small $k_3$ strengthens repeated key terms in short units (beneficial for decomposition), whereas large $k_3$ reduces this advantage and hurts long-document domains. The curve is smooth with a broad optimum $k_3 \in [0, 0.8]$, indicating stable tuning.

Table~\ref{tab:bright-descweight-nDCG@10} presents the complete experimental results in Figure~\ref{fig:desc-weight-single}. SBERT peaks near description weight $\lambda=0.5$, confirming that balancing the original sub-query and its interpretation yields the best alignment with document embeddings. Over-weighting interpretations $\lambda\ge 0.8$ or the sub-query $\lambda\le 0.2$ both reduce effectiveness, especially on theory-heavy sets.

\begin{table*}[!t]
  \centering
  \caption{Performance of dense retriever (SBERT) at varying interpretation weights.}
  \label{tab:bright-descweight-nDCG@10}
  %\small
  \setlength{\tabcolsep}{11pt}
  \begin{tabular}{%
  c| % desc_weight
  c % Avg.All (formerly Avg.)
  c| % Avg.ES (formerly StackExchange/Avg.)
  r@{\;\,}r@{\;\,}r@{\;\,}r@{\;\,}r@{\;\,}r@{\;\,}r
  r@{\;\,}r
  r@{\;\,}r@{\;\,}r}
  \toprule    
   \multirow{2}{*}{$\mathbf{interp._w}$}
   & \multirow{2}{*}{\textbf{Avg. all}}
   & \multirow{2}{*}{\textbf{Avg. SE}}
   & \multicolumn{7}{c}{\textit{StackExchange}}
   & \multicolumn{2}{c}{\textit{Coding}}
   & \multicolumn{3}{c}{\textit{Theorem‐based}} \\
  \cmidrule(lr){4-10} \cmidrule(lr){11-12} \cmidrule(lr){13-15}
   &
   &
   & Bio. & Earth. & Econ. & Psy. & Rob. & Stack. & Sus.
   & Leet. & Pony
   & AoPS & TheoQ. & TheoT. \\
    \midrule
    0.9 & 19.79 & 21.42 & 22.34 & 30.86 & 18.19 & 25.81 & 10.05 & 18.35 & 24.33 & 18.86 & 13.17 & 3.79 & 29.02 & 22.71 \\
    0.8 & 20.72 & 22.36 & 23.17 & 31.93 & 19.16 & 27.05 & 10.73 & 19.13 & 25.38 & 19.96 & 13.93 & 5.10 & 29.99 & 23.07 \\
    0.7 & 21.53 & 22.73 & 24.17 & 32.70 & 19.04 & 27.46 & 11.29 & 19.64 & 24.81 & 22.56 & 14.77 & 6.40 & 31.25 & 24.24 \\
    0.6 & 22.27 & 23.40 & 25.07 & 32.60 & 20.12 & 28.11 & 13.10 & 19.86 & 24.92 & 24.28 & 15.68 & 6.18 & 32.09 & 25.23 \\
    0.5 & \textbf{22.76} & 23.67 & 24.97 & 32.30 & 20.77 & 28.00 & 13.79 & 20.24 & 25.59 & 25.22 & 17.11 & 6.19 & 33.22 & 25.79 \\
    0.4 & 22.40 & 23.19 & 24.80 & 31.41 & 19.91 & 27.71 & 13.93 & 19.66 & 24.88 & 25.05 & 17.81 & 5.04 & 32.99 & 25.61 \\
    0.3 & 22.04 & 22.76 & 24.13 & 30.70 & 19.65 & 27.03 & 13.48 & 19.86 & 24.48 & 25.25 & 17.70 & 4.72 & 32.66 & 24.80 \\
    0.2 & 21.52 & 21.93 & 24.12 & 29.81 & 18.63 & 25.72 & 12.75 & 18.98 & 23.53 & 25.13 & 18.01 & 4.35 & 32.48 & 24.78 \\
    0.1 & 20.99 & 21.31 & 24.00 & 28.71 & 17.84 & 24.45 & 12.50 & 18.57 & 23.07 & 24.35 & 18.20 & 4.11 & 31.66 & 24.41 \\

    \bottomrule
  \end{tabular}
\end{table*}

\heading{Fine-tuning Paradigm}
Table~\ref{tab:joint vs two-stage full} presents the complete experimental results in Table~\ref{tab:joint_vs_twostage}. Two-stage training outperforms joint training for both BM25 and SBERT, with the largest gains on StackExchange and Theorem-based. Decoupling decomposition/inter\-pretation learning from retrieval scoring likely reduces optimization interference and improves stability.

\begin{table*}[!t]
  \centering
  \caption{nDCG@10 on BRIGHT: Joint vs. Two-Stage Training}
  \label{tab:joint vs two-stage full}
  \setlength{\tabcolsep}{10pt}
  \begin{tabular}{%
  l| % method
  c % Avg.All (formerly Avg.)
  c| % Avg.ES (formerly StackExchange/Avg.)
  r@{\;\,}r@{\;\,}r@{\;\,}r@{\;\,}r@{\;\,}r@{\;\,}r
  r@{\;\,}r
  r@{\;\,}r@{\;\,}r}
  \toprule    
   \multirow{2}{*}{\textbf{Method}}
   & \multirow{2}{*}{\textbf{Avg. all}}
   & \multirow{2}{*}{\textbf{Avg. SE}}
   & \multicolumn{7}{c}{\textit{StackExchange}}
   & \multicolumn{2}{c}{\textit{Coding}}
   & \multicolumn{3}{c}{\textit{Theorem‐based}} \\
  \cmidrule(lr){4-10} \cmidrule(lr){11-12} \cmidrule(lr){13-15}
   &
   &
   & Bio. & Earth. & Econ. & Psy. & Rob. & Stack. & Sus.
   & Leet. & Pony
   & AoPS & TheoQ. & TheoT. \\
    \midrule
    %\rowcolor{gray!20} 
    \multicolumn{15}{c}{\emph{Using BM25 Retriever}}\\
    \midrule   
    Joint                        & 28.3 & 35.4 & 46.7 & 49.9 & 27.2 & 38.4 & 24.5 & 33.8 & 27.0 & 23.6 & 6.9 & 5.5 & 28.3 & 28.1\\
    Two-Stage                    & 30.8 & 38.3 & 49.0 & 53.5 & 28.7 & 43.4 & 27.5 & 36.3 & 29.4 & 25.3 & 9.3 & 6.0 & 31.5 & 30.0\\
    \midrule
    %\rowcolor{gray!20} 
    \multicolumn{15}{c}{\emph{Using SBERT Retriever}}\\
    \midrule
    Joint                        & 20.8 & 21.8 & 22.7 & 29.9 & 18.9 & 25.7 & 12.0 & 18.9 & 24.8 & 24.0 & 14.8 & 5.1 & 30.2 & 22.1\\
    Two-Stage                    & 22.8 & 23.7 & 25.0 & 32.3 & 20.8 & 28.0 & 13.8 & 20.2 & 25.6 & 25.2 & 17.1 & 6.2 & 33.2 & 25.8\\ 
    \bottomrule
  \end{tabular}
\end{table*}

\heading{Fusion Methods}
Table~\ref{tab:fusion_method} presents the complete experimental results in Figure~\ref{fig:fusion_method}. Simple \textit{Sum} is consistently best. \textit{Max} discards complementary evidence and underperforms; \textit{RRF} helps some tails but lags on average; naive \textit{Concat} notably hurts dense retrieval (embedding dilution). These results support additive, facet-wise evidence aggregation for complex intents.

\begin{table*}[!t]
  \centering
  \caption{nDCG@10 on BRIGHT with different retrieval fusion methods.}
  \label{tab:fusion_method}
  \small
  \setlength{\tabcolsep}{13pt}
  \begin{tabular}{%
  l| % method
  c % Avg.All (formerly Avg.)
  c| % Avg.ES (formerly StackExchange/Avg.)
  r@{\;\,}r@{\;\,}r@{\;\,}r@{\;\,}r@{\;\,}r@{\;\,}r
  r@{\;\,}r
  r@{\;\,}r@{\;\,}r}
  \toprule    
   \multirow{2}{*}{\textbf{Method}}
   & \multirow{2}{*}{\textbf{Avg.All}}
   & \multirow{2}{*}{\textbf{Avg.SE}}
   & \multicolumn{7}{c}{\textit{StackExchange}}
   & \multicolumn{2}{c}{\textit{Coding}}
   & \multicolumn{3}{c}{\textit{Theorem‐based}} \\
  \cmidrule(lr){4-10} \cmidrule(lr){11-12} \cmidrule(lr){13-15}
   &
   &
   & Bio. & Earth. & Econ. & Psy. & Rob. & Stack. & Sus.
   & Leet. & Pony
   & AoPS & TheoQ. & TheoT. \\
    \midrule
    \multicolumn{15}{c}{\emph{Using BM25 Retriever}}\\
    \midrule   
    Sum                        & 30.8 & 38.3 & 49.0 & 53.5 & 28.7 & 43.4 & 27.5 & 36.3 & 29.4 & 25.3 & 9.3 & 6.0 & 31.5 & 30.0 \\
    Max                        & 24.8 & 29.6 & 37.6 & 40.6 & 22.0 & 32.5 & 20.8 & 31.3 & 22.5 & 20.1 & 11.3 & 3.5 & 27.2 & 28.2 \\
    RRF (rrf$_k=5$)            & 28.3 & 35.2 & 45.2 & 49.0 & 27.9 & 39.0 & 26.1 & 31.6 & 27.6 & 23.6 & 9.0 & 4.8 & 25.7 & 30.2 \\
    Concat                     & 29.2 & 36.8 & 50.4 & 53.7 & 26.0 & 40.2 & 26.6 & 33.2 & 27.8 & 21.1 & 8.9 & 6.4 & 28.4 & 27.4 \\

    \midrule
    \multicolumn{15}{c}{\emph{Using SBERT Retriever}}\\
    \midrule
    Sum                        & 22.8 & 23.7 & 25.0 & 32.3 & 20.8 & 28.0 & 13.8 & 20.2 & 25.6 & 25.2 & 17.1 & 6.2 & 33.2 & 25.8 \\
    Max                        & 19.1 & 18.9 & 23.2 & 24.1 & 15.7 & 24.1 & 8.4 & 14.7 & 22.3 & 17.7 & 22.8 & 3.5 & 28.7 & 24.2 \\
    RRF (rrf$_k=5$)            & 21.5 & 22.5 & 24.8 & 31.1 & 19.2 & 27.3 & 11.8 & 18.5 & 25.1 & 22.6 & 15.7 & 4.6 & 32.1 & 24.8 \\
    Concat                     & 12.4 & 12.4 & 11.9 & 22.5 & 9.4 & 19.3 & 5.1 & 8.3 & 9.9 & 17.0 & 8.8 & 2.1 & 17.8 & 16.8 \\ 
    \bottomrule
  \end{tabular}
\end{table*}

\heading{Reasoning-based Retriever of ReasonIR-8B.}
Table~\ref{tab:reasonir_results_full} presents the complete experimental results in Table~\ref{tab:reasonir_results}.
With ReasonIR-8B, systems built on single long-form rewrites lead on StackExchange and Theorem-based, while \textbf{ReDI} is competitive on Coding. This highlights a retriever-adaptive trade-off: specialized high-capacity dense encoders favor globally coherent rewrites; decomposition brings larger benefits with lexical or lighter dense retrievers and on modular tasks.

\begin{table*}[!t]
  \centering
  \caption{nDCG@10 on BRIGHT Benchmark with ReasonIR-8B retriever.}
  \label{tab:reasonir_results_full}
  %\small
  \setlength{\tabcolsep}{6pt}
  \begin{tabular}{%
  l % method
  l|% param
  c % Avg.All (formerly Avg.)
  c| % Avg.ES (formerly StackExchange/Avg.)
  r@{\;\,}r@{\;\,}r@{\;\,}r@{\;\,}r@{\;\,}r@{\;\,}r
  r@{\;\,}r
  r@{\;\,}r@{\;\,}r}
    \toprule
    \multirow{2}{*}{\textbf{QU Model}}
      & \multirow{2}{*}{\textbf{Params}}
      & \multirow{2}{*}{\textbf{Avg. all}}
      & \multirow{2}{*}{\textbf{Avg. SE}}
      & \multicolumn{7}{c}{\textit{StackExchange}}
      & \multicolumn{2}{c}{\textit{Coding}}
      & \multicolumn{3}{c}{\textit{Theorem‐based}}\\
    \cmidrule(lr){5-11} \cmidrule(lr){12-13} \cmidrule(lr){14-16}
      &
      & 
      &
      & Bio.\ & Earth.\ & Econ.\ & Psy.\ & Rob.\ & Stack.\ & Sus.
      & Leet.\ & Pony
      & AoPS & TheoQ.\ & TheoT. \\
    \midrule
    GPT-4    & -     & 28.8 & 31.2 & 42.6 & 40.9 & 30.0 & 36.7 & 19.7 & 23.2 & 25.3 & 30.2 & 19.6 & 7.9 & 33.7 & 37.9 \\
    TongSearch         & 7B    & 31.8 & 34.2 & 45.3 & 44.7 & 30.9 & 40.2 & 20.4 & 27.2 & 30.7 & 26.4 & 22.9 & 10.3 & 38.0 & 37.7 \\
    DIVER         & 14B   & 32.2 & 35.9 & 47.8 & 44.6 & 32.2 & 43.7 & 24.6 & 30.8 & 27.6 & 31.2 & 11.4 & 9.1 & 40.7 & 37.5 \\
    \textbf{ReDI}                          & 8B    & 30.7 & 32.7 & 41.8 & 43.9 & 26.5 & 37.4 & 21.3 & 32.2 & 25.5 & 35.7 & 21.3 & 8.6 & 39.0 & 34.7 \\
    \bottomrule
  \end{tabular}
  \medskip
\end{table*}

\end{document}